\documentclass[journal=jctcce,amsmath,amssymb,manuscript=article]{achemso}
\usepackage{multirow}
\usepackage{subfigure}
\usepackage{color}
\usepackage{epstopdf}
\usepackage[normalem]{ulem}
\usepackage{amsmath}
\usepackage{booktabs,siunitx,array}
\usepackage{longtable, booktabs}
\usepackage{ulem}
\usepackage{listings}
\usepackage{import}
\usepackage{xcolor}
\usepackage{framed}
\usepackage{mdwlist}
\usepackage{diagbox}
\usepackage{eufrak}
\usepackage{bm}
\usepackage{graphicx}
\usepackage{dcolumn}
\usepackage{algorithm}
\usepackage{etoolbox}
\usepackage{algpseudocode}
\usepackage{csquotes}
\usepackage[unicode]{hyperref}
\usepackage{longtable}
\usepackage{array}
\usepackage{subfigure}
\usepackage{siunitx}
\usepackage{amsmath}
\usepackage{color}
\usepackage{ifthen}
\usepackage{threeparttable}
\usepackage{float}
\title{Efficient structural relaxation based on the random phase approximation: Applications to the water clusters}

\author{Muhammad N. Tahir}
\affiliation{Institute of Physics, Chinese Academy of Sciences, Beijing 100190, China}
\author{Honghui Shang}
       \affiliation{University of Science and Technology of China, Hefei 230026, China}
\email{shh@ustc.edu.cn}

\author{Jia Li}
\email{li.jia@sz.tsinghua.edu.cn}
\affiliation{Shenzhen Geim Graphene Center and Institute of Materials Research, Tsinghua Shenzhen International Graduate School, Tsinghua University, Shenzhen 518055, China}

\author{Xinguo Ren}
\email{renxg@iphy.ac.cn}
\affiliation{Institute of Physics, Chinese Academy of Sciences, Beijing 100190, China}

\begin{document}
\newcommand{\calI}{ {\mathcal{I}}}
\newcommand{\calJ}{ {\mathcal{J}}}
\newcommand{\calK}{ {\mathcal{K}}}
\newcommand{\calL}{ {\mathcal{L}}}
\newcommand{\calU}{ {\mathcal{U}}}
\newcommand{\calV}{ {\mathcal{V}}}
\newcommand{\kpe}{\mathbf{k}\!\cdot\!\mathbf{p}\,}
\newcommand{\Kpe}{\mathbf{K}\!\cdot\!\mathbf{p}\,}
\newcommand{\bfr}{ {\bf r}} % definitions of r as a vector
\newcommand{\bfrp}{ {\bf r'}} % definitions of r' as a vector
\newcommand{\tp}{ {t^\prime}} % definitions of t'
\newcommand{\bfrpp}{ {\bf r^{\prime\prime}}} % definitions of r' as a vector
\newcommand{\bfR}{ {\bf R}} % definitions of R as a vector
\newcommand{\bfq}{ {\bf q}} % definitions of q as a vector
\newcommand{\bfp}{ {\bf p}} % definitions of q as a vector
\newcommand{\bfk}{ {\bf k}} % definitions of k as a vector
\newcommand{\bfG}{ {\bf G}} % definitions of k as a vector
\newcommand{\bfGp}{ {\bf G'}} % definitions of k as a vector
\newcommand{\dv}{ {\Delta \hat{v}}} % definitions of k as a vector
\newcommand{\sigmap}{\sigma^\prime} % definitions of k as a vector
\newcommand{\omegap}{\omega^\prime} % definitions of k as a vector
\newcommand{\omegapp}{\omega^{\prime\prime}} % definitions of k as a vector
\newcommand{\bracketm}[1]{\ensuremath{\langle #1   \rangle}}
\newcommand{\bracketw}[2]{\ensuremath{\langle #1 | #2  \rangle}}
\newcommand{\bracket}[3]{\ensuremath{\langle #1 | #2 | #3 \rangle}}
\newcommand{\ket}[1]{\ensuremath{| #1 \rangle}}
\newcommand{\GnWn}{\ensuremath{G_0W_0}\,}
\newcommand{\tn}[1]{\textnormal{#1}}
\newcommand{\f}[1]{\footnotemark[#1]}
\newcommand{\mc}[2]{\multicolumn{1}{#1}{#2}}
\newcommand{\mcs}[3]{\multicolumn{#1}{#2}{#3}}
\newcommand{\mcc}[1]{\multicolumn{1}{c}{#1}}
\newcommand{\refeq}[1]{(\ref{#1})} % equation reference, sets the number in brackets
\newcommand{\refcite}[1]{Ref.~\cite{#1}} % equation reference, sets the number in brackets
\newcommand{\refsec}[1]{Sec.~\ref{#1}} % reference to sections
\newcommand\opd{d}
\newcommand\im{i}
\def\bra#1{\mathinner{\langle{#1}|}}
\def\ket#1{\mathinner{|{#1}\rangle}}
\newcommand{\braket}[2]{\langle #1|#2\rangle}
\def\Bra#1{\left<#1\right|}
\def\Ket#1{\left|#1\right>}

\newcommand{\fatr}{\mathbf{r}}

\newcommand{\XR}[1]{{\color{blue}{\bf #1 }}}

\newcommand{\Or}{\mathcal{O}}
\newcommand{\ie}{\textit{i.e.}{}}
\renewcommand{\Im}{\mathrm{Im}~}
\newcommand{\Tr}{\mathrm{Tr}}

\newlength\replength
\newcommand\repfrac{.33}
\newcommand\dashfrac[1]{\renewcommand\repfrac{#1}}
\setlength\replength{8.5pt}
\newcommand\rulewidth{.6pt}
\def\NoNumber#1{{\def\alglinenumber##1{}\State #1}\addtocounter{ALG@line}{-1}}
\begin{tocentry}
	\centering
	\includegraphics[width=0.6\textwidth]{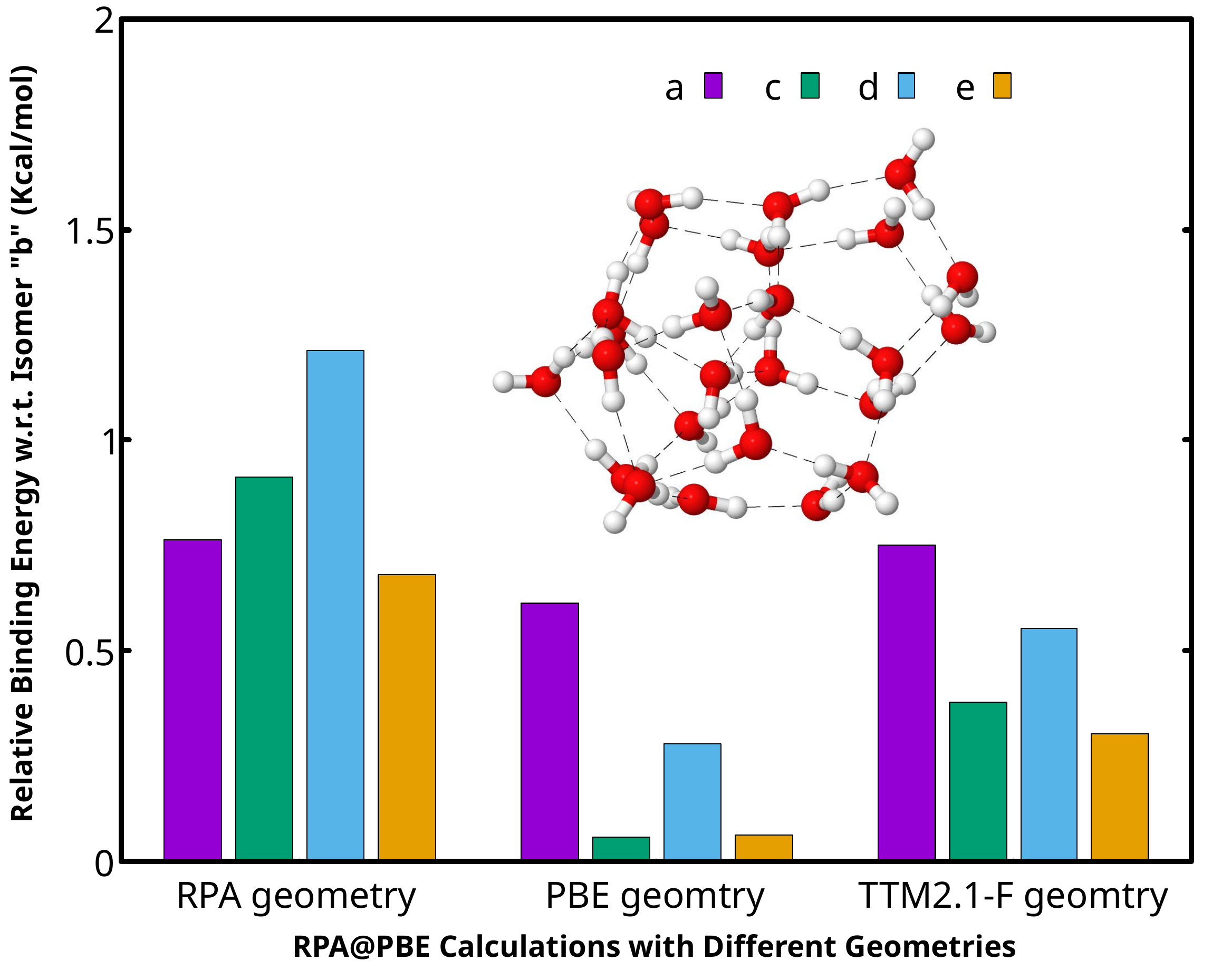} 
\end{tocentry}

\begin{abstract}
We report an improved implementation for evaluating the analytical gradients of the random phase approximation (RPA) electron-correlation
energy based on atomic orbitals and the localized resolution of identity scheme.
The more efficient RPA force calculations allow us to relax structures of medium-size water clusters. Particular attention is paid to 
the structures and energy orderings of the low-energy isomers of (H$_2$O)$_n$ clusters with $n=21$, 22, and 25.  It is found that the energy ordering
of the low-energy isomers of these water clusters are rather sensitive to how their structures are determined. 
For the five low-energy isomers of  (H$_2$O)$_{25}$, the RPA energy ordering based on the RPA geometries is quite different from that based on
the geometries relaxed by lower-level theories, in contrast with the situation of small water clusters like the water hexamer. 
The standard RPA underbinds the water clusters, and this underbinding behavior gets more pronounced as the complete basis set (CBS) limit is approached. The renormalized single excitation (rSE) correction remedies this underbinding, giving rise to a noticeable overbinding behavior at finite
basis sets. However, as the CBS limit is approached, RPA+rSE yields an accuracy for the binding energies that is comparable to the best
available double hybrid functionals, as demonstrated for the WATER27 testset.

%(i) For energetically very close isomers, the renormalized single excitation  (rSE) correction to RPA can change the energy ordering.
%(ii) The high sensitivity of the energy hierarchy to the underlying water cluster structures.\
%(iii) The importance of the non-local electron correlations. 
\end{abstract}

\section{Introduction}
%\XR{I think the focus of the introduction should be on RPA force and why it is important to be able to handle large systems, instead of water clusters. Water clusters are the benchmark systems, and should be briefly discussed, but not the focus.}
Water clusters (H$_2$O)$_n$ are the building blocks of bulk water -- the most important substance for life on earth. Studying the structures 
and properties of water clusters is a crucial step towards a molecular-level understanding of water in its condensed phases, including both liquid water
and ice \cite{Ludwig:2001,Brini/etal:2017}. Moreover, water clusters themselves are important components of biological systems and
are abundant in atmosphere, and hence studying the properties of water clusters is of high scientific interest by its own. 
Experimental and computational studies of water clusters have been continued for decades, yet numerous questions remain to be addressed, in particular regarding the energy rankings of low-lying isomers at each cluster size \cite{Bulusu/etal:2006}. 
The commonly used computational approaches for these studies range from classical force fields to 
density functional approximations (DFAs) and \textit{ab initio} quantum chemistry methods. 
%Identifying the global mininum and low-energy structures of small water clusters are of key interest. 
Due to the rapid increase  in the numbers of 
%of the amount of 
local minimum structures of (H$_2$O)$_n$ as the size $n$ grows, global structure search has 
mainly been carried out based on classical force fields, generated either empirically \cite{Jorgensen/etal:1983} or fitted to \textit{ab initio} data \cite{Fanourgakis/Xantheas:2008,Babin/etal:2012}.
The initially determined structures and their energies can then be refined using more accurate first-principles methods \cite{Bulusu/etal:2006}. %On the endeavor to fully understand the property of water clusters and eventually bulk water, it is of crucial importance to have inexpensive yet sufficiently accurate quantum chemical methods that can treat electrostatics, hydrogen bonding, and van der Walls (vdW) interactions on the same footing. 
In the quest for fully understanding the properties of water clusters (and ultimately bulk water), it is crucial to have inexpensive yet sufficiently accurate quantum chemical methods that can treat electrostatics, hydrogen bonding and van der Walls (vdW) interactions on an equal footing.
In this regard, the second-order M{\o}ller-Plesset perturbation theory (MP2) \cite{Moller/Plesset:1934}, simplest post-Hartree-Fock quantum-chemistry method,  
plays a pivotal role in the study of water clusters as it often delivers highly reliable results, and is often used to benchmark the accuracy of 
DFAs \cite{Santra/etal:2007,Biswajit/etal:2008}.

In recent years, the random phase approximation (RPA) \cite{Bohm/Pines:1953, Furche:2001, Ren/etal:2012b, Eshuis/Bates/Furche:2012,Hesselmann/Goerling:2011}, formulated
as an orbital-dependent fifth-rung functional of density functional theory (DFT) \cite{Langreth/Perdew:1977,Gunnarsson/Lundqvist:1976,Perdew/Schmidt:2001}, has emerged
as a powerful first-principles approach to evaluate the electronic ground-state energies of molecules and extended materials. 
Compared to lower-rung functionals, one prominent advantage of RPA is that it captures seamlessly the long-range van der Waals (vdW) interactions \cite{Dobson_2005, Dobson/Could:2012}.
%Benchmark calculations showed that RPA-based methods perform well for describing the interaction strength between weakly bonded molecules \cite{Toulouse/etal:2009,Zhu/etal:2010,Hesselmann/Goerling:2011b,Ren/etal:2011,Eshuis/Furche:2011,Ren/etal:2013} and are particularly suitable for capturing delicate energy differences between different structural configurations \cite{Ren/etal:2009,Schimka/etal:2010} and  crystalline polymorphs \cite{}.
Benchmark calculations showed that
RPA-based methods can accurately describe the interactions within weakly bonded molecular complexes \cite{Toulouse/etal:2009,Zhu/etal:2010,Hesselmann/Goerling:2011b,Ren/etal:2011,Eshuis/Furche:2011,Ren/etal:2013} and are capable of capturing 
the delicate energy differences between different structural configurations \cite{Ren/etal:2009,Schimka/etal:2010} and crystalline polymorphs  \cite{Lebegue/etal:2010,Zhang/Cui/Jiang:2018,Sengupta/Bates/Ruzsinszky:2018,Cazorla/Gould:2019,Yang/Ren:2022}. Furthermore,
compared to MP2, RPA can describe the polarization effects beyond the second order \cite{Dobson:1994} and is applicable to small-gap and metallic
systems. As such, there is a strong interest in applying RPA-based methods to water systems. Preliminary results on
water clusters \cite{Chedid/Jocelyn/Eshuis:2021,Muhammad_2022}, liquid water \cite{DelBen/etal:2015,Yao/Kanai:2021}, 
and ices \cite{Macher/etal:2014} show the great promise of
RPA-based methods in characterizing small energy differences in water systems. 
Given such initial successes, it is natural to think of determining the structures of water clusters based on the PRA
methods. Previously, MP2 is often employed as the high-level benchmark \textit{ab-initio} approach to determine the structures of water clusters
\cite{Bulusu/etal:2006,Biswajit/etal:2008}. In this regard, one expects that RPA may provide a competitive alternative.

%is most commonly used to describe the excitation of many-body systems and has gained popularity in %appealing use during
%the last two decades in its formulation as a density functional approximation. Therefore, the RPA forms a connection between
%density functional theory (DFT) and many-body methods as it has its roots in many-body
%theories such as Green’s function theory\cite{Blaizot/Ripka:1986,Luttinger/Ward:1961} or the coupled
%cluster theory\cite{Cizek_Paldus_1969,doi:10.1002/qua.560260825,doi:10.1063/1.465552, doi:10.1063/1.468022} and in DFT. The RPA represents a sophisticated functional, obtained when coupling
%the adiabatic connection provides the correlation energy which depends on the single particle orbitals and orbital energies.   The alternative to DFT, the wave function-based method such as coupled cluster methods or second-order M\o{}ller-Plesset (MP2) perturbation theory\cite{Moller/Plesset:1934} are computationally more complex.\\

%The RPA as a post-Kohn-Sham (KS) functional has gained popularity due to its good description of dispersion\cite{Dobson_2005, Dobson/Could:2012} and the ability to describe the system with small-gap (e.g., predicting the minima of d- and f-shell transition metal compounds)\cite{Burow_2014}, where other methods, e.g., MP2,  drastically fail. RPA is a non-empirical, robust, and relatively inexpensive method that has many desirable features of many purpose methods that can be applied to a wide range of different systems under different chemical environments.  
To determine the molecular structures 
% \textcolor{red}{(atomic structure deals with electron, proton and neutron. there should be molecular structure)} 
at the RPA level,  the analytical gradients of the RPA energy with respect to the atomic displacements need to be
evaluated. Within the last decade, much effort has been devoted to implementing the analytical gradients of RPA, mostly for molecular geometries. Among these,  Rekkedal \textit{et al.} \cite{Rekkedal/etal:2013} were the first to develop an $\mathcal{O}$($N^{6}$)-scaling algorithm for evaluating the analytical gradient of RPA on top of the Hartree-Fock (HF) reference, utilizing the ring-coupled-cluster-doubles (rCCD) formulation of RPA. Shortly after, 
a similar rCCD-based analytical gradient formalism for RPA was developed by 
 Mussard \textit{et al.} \cite{Mussard_2014} within the range-separation framework. Based on the resolution of identity (RI) 
 technique and Lagrangian formalism, Burow \textit{et al.} \cite{Burow_2014} reduced the scaling behavior of RPA gradient calculations to $\mathcal{O}$($N^{4}\text{log}(N)$), taking the size dependence of the imaginary frequency grid into account. Further development was carried out by  Ramberger \textit{et al.} \cite{Ramberger/etal:2017} who 
 achieved an $\mathcal{O}(N^3)$-scaling RPA gradient algorithm for periodic systems based on the Green's function formalism and plane-wave
 basis set. The $\mathcal{O}(N^3)$ scaling was enabled by the space-time algorithm for evaluating the $GW$ self-energy \cite{Rojas/Godby/Needs:1995,Liu/etal:2016}, which is needed
 in this particular RPA gradient formalism \cite{Ramberger/etal:2017}.
 This algorithm was further reduced to $\mathcal{O}$($N^{2}$) by Beuerle \textit{et al.} \cite{Beuerle/Ochsenfeld:2018} in terms of an  
 atomic-orbital (AO)
 formulation whereby the sparsity of the AO-based integrals can be exploited. Recently, the present authors also developed a RPA gradient approach \cite{Muhammad_2022} for molecular calculations within the localized RI (LRI) scheme \cite{Ihrig/etal:2015}, which works for both Gaussian-type orbital (GTO) and numerical atomic
 orbital (NAO) basis sets. In the present work, we further improve our algorithm and code, achieving a \textit{de facto} $\mathcal{O}(N^3)$-scaling for RPA gradient calculations. This enables the structure relaxation of molecules of much bigger size at the RPA level, especially for water clusters of
 medium size ($n>20)$.
 %$\mathcal{O}$($N^{4}$) scaling by Muhammad \textit{et al.}\cite{Muhammad_2022} based on resolution of identity (RI) and with DFT references rather than HF. The analytical gradient of RPA for extended system is implemented by has $\mathcal{O}$($N^{3}$) scaling with HF reference. \\

RPA has been previously used to study small water clusters. An earlier study of water clusters using RPA in a dual hybrid form (
the so-called dRPA75 functional \cite{Mezei/etal:2015}) was carried out by Mezei \textit{et al.} \cite{Mezei_2016}. Using the standard RPA, 
Chedid \textit{et al.} performed a systematic study
of the WATER27 \cite{WATER27} testset, benchmarking the performance of RPA for the structural, energetic, and
vibrational properties of this testset. They particularly studied the basis set dependence of the RPA binding energies.
In consistent with previous experience,
they found a substantial underbinding and pronounced basis set dependence of RPA for water clusters. However, despite its overall underbinding
behavior, RPA does yield the correct energy ordering for the low-lying isomers of 
water hexamers \cite{Chedid/Jocelyn/Eshuis:2021,Muhammad_2019}. Previously, it
was found that the underbinding behavior of RPA can be largely cured by adding the renormalized single excitation (rSE)
corrections \cite{Ren/etal:2013,Klimes/etal:2015}, and the efficacy of rSE was also observed in water-cluster
systems \cite{Muhammad_2022,Modrzejewski/etal:2021}. Thus, we expect that adding rSE corrections to RPA is essential
for the RPA-based methods to describe the water systems quantitatively. 

In this work, after recapitulating the basic algorithm of our implementation, we present a performance study of our RPA gradient implementation,
thereby checking its scaling behavior, parallel efficiency, and numerical accuracy. Then we look at the energy hierarchy of the low-lying isomers of
water clusters of larger size, with $n=21$, 22, and $25$. In particular, we check how the energy ordering of different water isomers change
with the level of theories employed to relax their structures. We found that, unlike the water hexamers where going from PBE to RPA geometries the
energy ordering is preserved, for larger clusters like
(H$_2$O)$_{25}$, this is not the case any more. If one trusts that RPA-based methods yield 
reliable energy hierarchy of water clusters, it seems that, in general, it is also important to determine the geometries 
consistently at the RPA level. We then also checked the basis set convergence for both RPA and RPA+rSE. It is found that as one increases
the basis size towards the complete basis set (CBS) limit, the RPA becomes even more underbinding for water clusters, in consistent with
Ref.~\citenum{Chedid/Jocelyn/Eshuis:2021}, whereas the RPA+rSE evolves from 
an overbinding behavior towards an excellent agreement with the most accurate results yielded by double hybrid functionals, 
reported in the literature \cite{}. Hopefully the present study will shed new light on the application of RPA-based methods to 
the study of water systems.

\section{\label{sec:theory}Theory and Algorithm}
%The random phase approximation is the method to calculate the correlation energy from excited configurations in the post-SCF manner like second-order M{\o}ller--Plesset perturbation theory (MP2) and configuration interaction method.  The RPA is a way to renormalize the second-order perturbation theory in a slightly different way than MP2. The RPA in its two flavors particle-hole RPA and particle-particle RPA has been extensively studied for non-local electron correlation energy demonstrated for molecules \cite{Furche:2001,Fuchs/Gonze:2002,Toulouse/etal:2009,Zhu/etal:2010,Hesselmann/Goerling:2011b,Ren/etal:2011,Eshuis/Furche:2011, Muhammad_2019, Muhammad_2022,Aggelen/etal:2013,Peng/Steinmann/etal:2013,Yang/Aggelen/Steinmann/etal:2013}, solids \cite{Harl/Kresse:2008,Harl/Kresse:2009,Harl/Schimka/Kresse:2010,Lebegue/etal:2010,Nguyen/deGironcoli:2009,Lu/Li/Rocca/Galli:2009,Casadei/etal:2012,Casadei/etal:2016}, surfaces \cite{Ren/etal:2009,Schimka/etal:2010}, interfaces \cite{Mittendorfer/etal:2011,Olsen/etal:2011}, 
%and defects \cite{Bruneval:2012,Kaltak/Klimes/Kresse:2014,Kaltak/Klimes/Kresse:2014b}. Here, in the next, we will provide the expression of direct particle-hole RPA and simply refer to it as RPA. 

\subsection{RPA total energy within the LRI approximation}
\label{Sec:RPA_ene_LRI}

%\subsection{RPA gradients within the RI framework}

% \subsection{Atom pairs and the derivatives of the expansion coefficients within the LRI scheme. }
We start with the RPA total-energy expression and then consider its derivatives with respect to the atomic displacements, highlighting the
reduction of the computational cost brought about by the localized RI (LRI). The RPA total energy is given by
\begin{equation}
    E^{\text{RPA}}=E^{\text{HF}} +E^{\text{RPA}}_{c}
    \label{eq:RPA_etot}
\end{equation}
where $E^{\text{HF}}$ is the Hartree-Fock total energy evaluated using orbitals generated from a preceding calculation based on
density functional approximations (DFAs).
The Perdew-Burke-Ernzerhof (PBE) generalized gradient approximation (GGA) is often used to generate the orbitals and orbital energies for
RPA calculations.  The
key component in Eq.~(\ref{eq:RPA_etot}),  the RPA correlation energy, is formally given by
\begin{equation}
    E^{\text{RPA}}_{c}=\dfrac{1}{2\pi} \int _{0}^{\infty}d\omega \text{Tr} \Big[ \text{ln} \big( 1-\chi^{0}(i\omega)v\big) +\chi^{0}(i\omega)v \Big]
    \label{eq:E_c_RPA}
\end{equation}
where ${\chi}^0$ is non-interacting Kohn-Sham (KS) response function, which has an explicit ``sum-over-states" expression,
\begin{equation}
{\chi}^0(\textbf{r},\textbf{r}^{\prime},i\omega) = \sum_{m} \sum_{n} \frac{(f_m-f_n)\psi_m(\bfr)\psi_n(\bfr)\psi_n(\bfrp)\psi_m(\bfrp)}{\epsilon_m - \epsilon_n -i\omega}\, ,
\label{eq:chi0_realspace}
\end{equation}
with $\psi_n$, $\epsilon_n$, and $0\le f_n \le 2$ being the KS orbitals, their energies and occupation numbers. %Throughout this paper for further discussion, we have used $m$ as occupied and $n$ as the unoccupied orbitals. 
Computationally, ${\chi}^0$ and ${v}$ in Eq.~(\ref{eq:E_c_RPA}) can be interpreted as the matrix form of the non-interacting
density response function and the bare Coulomb interaction, represented in terms of a set of suitable basis functions. 
For simplicity, here closed-shell systems (hence $f_n =0$ or 2) and real orbitals are assumed. Extensions to spin-collinear cases and complex orbitals are straightforward.

Within an atomic-orbital (AO) basis set framework, the KS molecular orbitals (MOs) are expanded in terms of a set of 
atom-centered basis functions $\{\varphi_{i,{\cal I}}(\bfr)\}$,
\begin{equation}
    \psi_n(\bfr) = \sum_{i,\calI} c_{i(\calI)}^n \varphi_{i,\calI}(\bfr-\bfR_{\calI})
\end{equation}
with $c_{i(\calI)}^n$ being the KS eigenvectors and $\bfR_\calI$ the position of the atom $\calI$, on which the basis function $i$ is centering. 
For clarity, here we explicitly indicate the atom on which each atomic basis function belongs to, and the summation over $i$ goes over
only those AOs centering on the atom $\calI$. 
Furthermore, the exact-exchange (EX) and RPA correlation energies are computed based on the RI approximation. Within this approximation, 
a set of 
auxiliary basis functions (ABFs) $\{P_\mu(\bfr)\}$ are employed to expand the products of two AOs, 
\begin{equation}
    \varphi_{i,\calI}(\bfr-\bfR_\calI)\varphi_{j,\calJ}(\bfr-\bfR_\calJ)=\sum_{\mu,\calU}C_{i(\calI),j(\calJ)}^{\mu(\calU)} P_\mu(\bfr-\bfR_\calU)
    \label{eq:RI_AO_expansion}
\end{equation}
where  $\bfR_\calU$ is the position of the atom on which the ABF $\mu$ is centered, and $C_{i(\calI),j(\calJ)}^{\mu(\calU)}$ are the expansion coefficients, 
with the atoms to which the AOs and ABFs belong explicitly indicated in parenthesis. 

In the global RI approximation \cite{Dunlap/Connolly/Sabin:1979,Whitten:1973,Feyereisen/Fitzgerald/Komornicki:1993,Ren/etal:2012}, 
the atom $\calU$ could be a third atom other than the atoms $\calI$ and $\calJ$;
however, in the LRI approximation \cite{Ihrig/etal:2015,Levchenko/etal:2015,Lin/Ren/He:2020} adopted in the present work, 
$\calU$ has to be either $\calI$ or $\calJ$, i.e.,
\begin{equation}
    \varphi_{i,\calI}(\bfr-\bfR_\calI)\varphi_{j,\calJ}(\bfr-\bfR_\calJ)=\sum_{\mu \in \calI }C_{i(\calI),j(\calJ)}^{\mu(\calI)} P_\mu(\bfr-\bfR_\calI) +
    \sum_{\mu \in \calJ }C_{i(\calI),j(\calJ)}^{\mu(\calJ)} P_\mu(\bfr-\bfR_\calJ)
    \label{eq:RI_AO_expansion_LRI}
\end{equation}
where ${\mu \in \calI }$ means the summation over ABFs $\{\mu\}$ are restricted to those centering on the atom $\calI$.
In the RI (LRI) formalism, another key quantity is the Coulomb matrix, which is the representation of the Coulomb operator in terms of ABFs,
\begin{equation}
    V_{\mu(\calU),\nu(\calV)} = \int d\bfr d\bfrp \frac{P_\mu(\bfr-\bfR_\calU)P_\nu(\bfrp-\bfR_\calV)}{|\bfr-\bfrp|}\, .
\end{equation}
Introducing 
%\begin{equation}
%    Q_{ij}^\mu =\sum_\nu C_{i,j}^\nu \left[V^{\frac{1}{2}}\right]_{\nu,\mu}\,
%\end{equation}
%and
\begin{equation}
    O^{\mu(\calU)}_{mn}=\sum_{i,\calI}\sum_{j,\calJ}\sum_{\nu,\calV} c_{i(\calI)}^m C^{\nu(\calV)}_{i(\calI),j(\calJ)} c_{j(\calJ)}^n
    \begin{bmatrix}
        V^{\frac{1}{2}}
    \end{bmatrix}_{\nu(\calV), \mu(\calU)}\, 
    \label{eq:O_integrals}
\end{equation}
where $V^{1/2}$ is the square root of the global Coulomb matrix,
then it is straightforward to show that the exact-exchange energy is given by
\begin{equation}
E_{x}^{\text{EX}}=-\sum_{m}^{occ}\sum_{n}^{occ} \sum_{\mu, \calU}O^{\mu(\calU)}_{mn} O^{\mu(\calU)}_{nm} \, .
\label{eq:EX_RI}
\end{equation} 
Further denoting
\begin{equation}
   \mathbf{\Pi}_{\mu(\calU),\nu(\calV)} (i\omega)=2\sum_{m}^{occ} \sum_{n}^{unocc}\frac{2(\epsilon_m-\epsilon_n)O_{mn}^{\mu(\calU)} O_{nm}^{\nu(\calV)}}{(\epsilon_m-\epsilon_n)^{2}+\omega^{2}}\, 
   \label{eq:Pi_matrix}
\end{equation}
(assuming integer occupations), the RPA correlation energy is then given by
\begin{equation} 
E_c^\text{RPA} = \frac{1}{2\pi} \int_0^\infty d\omega \text{Tr}\left[\text{ln}\left(\textbf{1}-\mathbf{\Pi}(i\omega)\right) + \mathbf{\Pi}(i\omega) \right]\, .
\label{eq:EcRPA_RI}
\end{equation}
More detailed derivations of Eqs.~(\ref{eq:EX_RI}) and (\ref{eq:EcRPA_RI}), on which our RPA force implementation is based, are given in
Refs.~\cite{Ren/etal:2012,Ren/etal:2012b}.

From the above presentation, one can see that a key step in the RPA energy (and subsequently the RPA force) calculations is to evaluate
the $O^{\mu(\calU)}_{mn}$ tensor, defined in Eq.~(\ref{eq:O_integrals}). Previously, this was done rather straightforwardly in terms of matrix multiplications, without sufficiently exploiting the sparsity in $C^{\nu(\calV)}_{i(\calI),j(\calJ)}$. To facilitate a better parallelization
efficiency and reduce the memory cost, we now change the strategy to calculate $O^{\mu(\calU)}_{mn}$ by taking into account the sparsity of 
$C^{\nu(\calV)}_{i(\calI),j(\calJ)}$. 
%In this section, we will provide the complete implementation details regarding to RPA-force. Previous information about atom pairs 
%would remain the same\cite{Muhammad_2022}.  
%Previously, we calculate $O^{\mu}_{nm}$ in the following way
%\begin{equation}
%    O^{\mu}_{nm}=\sum_{i,\calI}\sum_{j,J}\sum_{\nu \in I,J}c_{j,J}^{n}C^{\nu(V)}_{j(\calJ),i(\calI)}c_{i,\calI}^{m} 
%    \begin{bmatrix}
%        V^{\frac{1}{2}}
%    \end{bmatrix}_{\nu, \mu}
%\end{equation}
%If we restrict $I\geq J$, then we can write the above expression as 
%\begin{equation}
%   \begin{split}
%    O^{\mu}_{nm}&=\sum_{i,\calI}\sum_{j,J}\sum_{\nu \in I,J} \Bigg[ c_{j,J}^{n}C^{\nu(V)}_{j(\calJ),i(\calI)}c_{i,\calI}^{m} +c_{i(\calI)}^nC^{\nu(V)}_{j(\calJ),i(\calI)}c_{j,J}^{m}\Bigg]
%    \begin{bmatrix}
%        V^{\frac{1}{2}}
%    \end{bmatrix}_{\nu, \mu} \\&
%    =\sum_{i,\calI}\sum_{j,J}\sum_{\nu \in I,J} \Bigg[ c_{j,J}^{n}c_{i,\calI}^{m} +c_{i(\calI)}^nc_{j,J}^{m}\Bigg] C^{\nu(V)}_{j(\calJ),i(\calI)}
%    \begin{bmatrix}
%        V^{\frac{1}{2}}
%    \end{bmatrix}_{\nu, \mu}
%    \end{split}
%\end{equation}
%For massively parallel implementation we have changed the strategy to calculate $O^{\mu}_{ij}$ by multiplying the $C^{\mu}_{ij}$ with $c_{im}$ for each local pair of atoms. The local multiplication of $C^{\mu}_{ij}$ with $c_{im}$ is given by  
Specifically, using the property that $C^{\nu(\calV)}_{i(\calI),j(\calJ)}$ is nonzero only if $\calV=\calI$ or $\calV=\calJ$ within the 
LRI, we first calculate an intermediate quantity, given by local multiplication of  $C^{\nu(\calV)}_{i(\calI),j(\calJ)}$ with 
the KS eigenvectors, i.e.,
\begin{equation} \label{Eq:C_times_KSvec}
\begin{split}
S^{\nu(\calI)}_{i(\calI), m}  = \sum_{j \in \calJ} C^{\nu(\calI)}_{i(\calI), j(\calJ)} c_{j(\calJ)}^m \\
S^{\nu(\calI)}_{j(\calJ), m}  = \sum_{i \in \calI} C^{\nu(\calI)}_{j(\calJ),i(\calI)} c_{i(\calI)}^m 
\end{split}
\end{equation}
for each pair of atoms $\langle \calI, \calJ \rangle$ with $\calI \ge \calJ$ and $\calV=\calI$.
Similarly, we can calculate $S^{\nu(\calJ)}_{i(\calI), m}$ and $S^{\nu(\calJ)}_{j(\calJ), m}$ for $\calV=\calJ$.

%\begin{equation}
%\begin{bmatrix}
%    S^{\mu(\calI)}_{i(\calI), m} \\
%    S^{\mu(\calI)}_{j(\calJ), m}
%\end{bmatrix}
%    =
%    \begin{bmatrix} 
%     0&& C^{\mu(\calI)}_{i(\calI), j(\calJ)} \\
%     C^{\mu(\calI)}_{j(\calJ),i(\calI)} && 0
%    \end{bmatrix}
%    \begin{bmatrix}
%      c_{i(\calI)}^m \\
%      c_{j(\calJ)}^m
%    \end{bmatrix}
%\end{equation}
%for $\calV = \calI$. Similarly,
%for $\calV = \calJ$, we have
%\begin{equation}
%\begin{bmatrix}
%    S^{\mu(\calJ)}_{i(\calI),m} \\
%    S^{\mu(\calJ)}_{j(\calJ),m}
%\end{bmatrix}
%    =
%    \begin{bmatrix} 
%     0&& C^{\mu(J)}_{i(\calI), j(\calJ)} \\
%     C^{\mu(J)}_{j(\calJ),i(\calI)} && 0
%    \end{bmatrix}
%    \begin{bmatrix}
%      c_{i(\calI)}^m \\
%      c_{j(\calJ)}^m
%    \end{bmatrix}
%    \quad\text{.}
%\end{equation}
%In both cases, we can restrict that $\calI > \calJ$. In the special case of $\calI = \calJ$, we have
%\begin{equation}
%\begin{bmatrix}
%    S^{\mu(\calI)}_{i(\calI), m} 
%\end{bmatrix}
%    =
%    \begin{bmatrix} 
%     C^{\mu(\calI)}_{i(\calI), i(\calI)}  
%    \end{bmatrix}
%    \begin{bmatrix}
%      c_{i(\calI)}^m
%    \end{bmatrix}
%    \quad\text{.} 
%\end{equation}

With the intermediate quantity obtained, the target $O$-tensor can be calculated as  
\begin{equation}
    O^{\mu(\calU)}_{nm}=\sum_{\calJ \calV }\sum_{j \in \calJ, \nu \in \calV}c_{j(\calJ)}^n S^{\nu(\calV)}_{j(\calJ),m} 
    \begin{bmatrix}
        V^{\frac{1}{2}}
    \end{bmatrix}_{\nu(\calV), \mu(\calU)} \label{eq:O_integrals_1}
\end{equation}
This step helps a lot to deal with the storage of the three-index $C$ tensor and its transformation to MO representations.
%and is computationally efficient. 
%\XR{From the discussion so far, it is still not clear how this is achieved. Evaluating Eq.~(13) is still expensive and memory intensive.}

\subsection{RPA Forces within the LRI Framework} 
\label{Sec:force_with_LRI}
The force that an atom feels due to the RPA correlation energy is the negative of the gradient of RPA correlation energy, i.e.,
\begin{eqnarray}  \label{eq:EcRPA_force1}
	F^{\text{RPA}}_{c,A}= -\frac{dE^{\text{RPA}}_{c}}{d\mathbf{R}_{A}} &=& 	- \big\langle \dfrac{d E^{\text{RPA}}_{c}}{dC} \dfrac{dC}{d\mathbf{R}_{A}}\big \rangle - \big\langle \dfrac{d E^{\text{RPA}}_{c}}{dV} \dfrac{dV}{d\mathbf{R}_{A}} \big \rangle -\big\langle \dfrac{d E^{\text{RPA}}_{c}}{dc} \dfrac{dc}{d\mathbf{R}_{A}} \big \rangle - 
	\big\langle \dfrac{d E^{\text{RPA}}_{c}}{d\epsilon} \dfrac{d\epsilon}{d\mathbf{R}_{A}} \big \rangle  
  \nonumber \\
	&=&  \big\langle \Gamma^{(1)} \dfrac{dC}{d\mathbf{R}_{A}}\big \rangle + \big\langle \Gamma^{(2)} \dfrac{dV}{d\mathbf{R}_{A}} \big \rangle +
	\big\langle \Gamma^{(3)} \dfrac{dc}{d\mathbf{R}_{A}} \big \rangle + 
	\big\langle \Gamma^{(4)} \dfrac{d\epsilon}{d\mathbf{R}_{A}} \big \rangle  \, ,
\end{eqnarray}
where the first two terms depend on the atomic positions explicitly, while the last two terms, consisting of the analytical gradients of KS 
eigenvectors and eigenenergies, depend on the atomic positions implicitly and are obtained via density-functional perturbation theory (DFPT) \cite{Shang/etal:2017}.
In practical calculations, we introduce two frequency-integrated intermediate quantities,
\begin{eqnarray}
    \Upsilon^{\mu(\calU)}_{mn}=\int_{0}^{\infty} \frac{d\omega}{2\pi} \times \Big ( 
 \sum_{\nu,\calV} \frac{f_{m}-f_{n}}{\epsilon_m-\epsilon_n -i\omega} O^{\nu(\calV)}_{mn} \mathbf{W}^c_{\nu(\calV),\mu(\calU)}(i\omega) \Big ) 
\end{eqnarray}
and
\begin{eqnarray}
    \tilde{\Upsilon}^{\mu(\calU)}_{mn}=\int_{0}^{\infty}  \frac{d\omega}{2\pi} (f_{m}-f_{n}) \times  \Big ( 
\sum_{\nu,\calV} \frac{2\left[\omega^{2}-(\epsilon_{m}-\epsilon_{n})^{2}\right]}{\left[(\epsilon_{m}-\epsilon_{n})^{2} +\omega^{2}\right]^{2}}O^{\nu(\calV)}_{mn} \mathbf{W}^{c}_{\nu(\calV), \mu(\calU)}(i\omega) \Big ) 
\end{eqnarray}
where 
\begin{equation}
    \mathbf{W}^c(i\omega) = \Bigg [ \dfrac{\mathbf{\Pi}(i\omega)}{\mathbf{1}-\mathbf{\Pi}(i\omega)} \Bigg]\, ,
\end{equation}
with $\mathbf{\Pi}$ defined in Eq.~(\ref{eq:Pi_matrix}).
Then, for the case of integer occupations, the $\Gamma^{(n)}$ tensors introduced in Eq.~(\ref{eq:EcRPA_force1}) can be evaluated as 
\begin{equation} \label{eq.gamma1}
\begin{split}
\Gamma^{(1)}_{i(\calI)j(\calJ),\mu(\calU)}= -\frac{dE^{\text{RPA}}_{c}}{dC^{\mu(\calU)}_{i(\calI),j(\calJ)}} =  \sum_{\nu,{\calV}} V^{\frac{1}{2}}_{\mu(\calU),\nu(\calV)} \sum_{m,n}c_{i,\calI}^{m} 
 \Upsilon^{\nu(\calV)}_{mn} c_{j,\calJ}^{n}  \, \\
=  \sum_{\nu,{\calV}} V^{\frac{1}{2}}_{\mu(\calU),\nu(\calV)} \sum_{m}^{occ}\sum_{n}^{unocc}\left[ c_{i,\calI}^{m} 
 \Upsilon^{\nu(\calV)}_{mn} c_{j,\calJ}^{n}  + c_{i,\calI}^{n} 
 \Upsilon^{\nu(\calV)}_{nm} c_{j,\calJ}^{m} \right]
\end{split}
\end{equation}

\begin{equation} \label{eq.gamma2}
\begin{split}
\Gamma^{(2)}_{\mu(\calU),\nu(\calV)}= -\frac{dE^{\text{RPA}}_{c}}{dV_{\mu(\calU),\nu(\calV)}} =  
\sum_{\mu',\calU'} \sum_{\nu',\calV'} V^{-\frac{1}{2}}_{\mu(\calU), \mu'(\calU')} \Big( \sum_{m}^{occ}\sum_{n}^{unocc}O^{\mu'(\calU')}_{mn}  
\Upsilon^{\nu'(\calV')}_{nm} \Big) V^{-\frac{1}{2}}_{\nu'(\calV'), \nu(\calV)}\, ,
\end{split}
\end{equation}

\begin{equation} \label{eq.gamma3}
\begin{split}
\Gamma^{(3)}_{i(\calI),m}= -\frac{dE^{\text{RPA}}_{c}}{dc_{i(\calI)}^m} = 2\sum_{n}\sum_{j,\calJ}\sum_{\nu,\calV} 
Q^{\nu(\calV)}_{i(\calI),j(\calJ)}c_{j(\calJ)}^{n}  \Upsilon^{\nu(\calV)}_{nm}
\, ,
\end{split}
\end{equation}
with \begin{equation}
    Q^{\nu(\calV)}_{i(\calI),j(\calJ)}=\sum_{\mu,\calU}C^{\mu(\calU)}_{i(\calI),j(\calJ)} 
    \begin{bmatrix}
        V^{\frac{1}{2}}
    \end{bmatrix}_{\mu(\calU), \nu(\calV)}
\end{equation}
and 
\begin{equation} \label{eq.gamma4}
\begin{split}
\Gamma^{(4)}_{m}= -\frac{dE^{\text{RPA}}_{c}}{d\epsilon_{m}} = \sum_{n}\sum_{\nu,\calV} O^{\nu(\calV)}_{mn} \tilde{\Upsilon}^{\nu(\calV)}_{mn} \, .
\end{split}
\end{equation}

% \begin{equation}
%    \Big\langle\dfrac{dE_{c}^{RPA}}{dV} \dfrac{dV}{d\textbf{R}_{A}}\Big\rangle= \Big\langle\Gamma^{(2)}_{c}\dfrac{dV}{d\textbf{R}_{A}}\Big\rangle
% \end{equation}
Further details of the derivation of these equations can be found in Ref.~\citenum{Muhammad_2022}, although we have used slightly different
notations here by explicitly specifying the atomic positions of the AOs and ABFs. For $N_{at}$ total atoms, there should be $N_{at}(N_{at}-1)/2$ atom pairs which constitute the upper/lower triangular part of a matrix.  For onsite atom pairs, the derivatives are zero and so does the force. \\
 
 Next, we briefly explain how the traces in Eq.~(\ref{eq:EcRPA_force1}) is performed. First, we note that the global trace operation can be 
 separated into a summation over atomic pairs, 
 \begin{equation}
 \big\langle AB\rangle = 2\sum_{\calI > \calJ} \big\langle AB\big \rangle_{\langle \calI, \calJ \rangle}
 \end{equation}
 and for each pair, we perform local multiplications over the orbital indices. In fact, this separation is most convenient for the first
 two terms in Eq.~\ref{eq:EcRPA_force1}, involving the derivatives of expansion coefficients $C$ and the Coulomb matrix $V$ with respect to
 the atomic positions. Taking the first term for example, for each atomic pair $\langle \calI, \calJ \rangle$,  the following local operations are
 performed, utilizing the sparse property of $C_{i(\calI),j(\calJ)}^{\mu(\calU)}$, 
 \begin{equation} \label{Eq:trace_first_term}
    \begin{split}
        \big\langle \Gamma^{(1)} \dfrac{dC}{d\mathbf{R}_{A}} \big \rangle_{\langle \calI,\calJ \rangle} = \sum_{\mu \in \calI} \sum_{i\in \calI, j \in \calJ} \Gamma^{(1)}_{i(\calI)j(\calJ),\mu(\calI)}\dfrac{dC^{\mu(\calI)}_{j(\calJ),i(\calI)}}{d\textbf{R}_{A}} + \sum_{\mu \in \calJ} \sum_{i\in \calI, j \in \calJ} \Gamma^{(1)}_{i(\calI)j(\calJ),\mu(\calJ)}\dfrac{dC^{\mu(\calJ)}_{j(\calJ),i(\calI)}}{d\textbf{R}_{A}} \, ,
    \end{split}
\end{equation}
which amounts to two separate contractions over small rank-3 tensors.

% In case, we use the symmetry in $
%    \dfrac{dC^{\mu I}_{iI,jJ}}{d\textbf{R}_{A}}=\dfrac{dC^{\mu I}_{jJ,iI}}{d\textbf{R}_{A}} $ allows us to perform the calculation for atom pairs  $I>J$. For atom pair choice $I>J$ we can write Eq.~(\ref{Eq:trace_first_term}) as 
%
%\begin{equation}
%    \begin{split}
%        &\sum_{\mu \in I} \sum_{i\in I, j \in J} \Gamma^{(1)}_{iIjJ,\mu I}\dfrac{dC^{\mu I}_{jJ,iI}}{d\textbf{R}_{A}} + \sum_{\mu \in I} \sum_{i\in I,j \in J} \Gamma^{(1)}_{jJiI,\mu I}\dfrac{dC^{\mu I}_{iI, jJ}}{d\textbf{R}_{A}} +\\&
%        \sum_{\mu \in J} \sum_{i\in I, j \in J} \Gamma^{(1)}_{iIjJ,\mu J}\dfrac{dC^{\mu J}_{jJ,iI}}{d\textbf{R}_{A}} + \sum_{\mu \in J} \sum_{i\in I,j \in J} \Gamma^{(1)}_{jJiI,\mu J}\dfrac{dC^{\mu J}_{iI, jJ}}{d\textbf{R}_{A}} \\&
%        =\sum_{\mu \in I} \sum_{i\in I, j \in J}\Bigg[\Gamma^{(1)}_{iIjJ,\mu I}+ \Gamma^{(1)}_{jJ iI, \mu I}\Bigg] \dfrac{dC^{\mu I}_{iI,jJ}}{d\textbf{R}_{A}} +\sum_{\mu \in J} \sum_{i\in I, j \in J}\Bigg[\Gamma^{(1)}_{iIjJ,\mu J}+ \Gamma^{(1)}_{jJ iI, \mu J}\Bigg] \dfrac{dC^{\mu J}_{iI,jJ}}{d\textbf{R}_{A}}
%    \end{split}
%\end{equation}

%\begin{equation}
%    \Gamma'^{(1)}_{iIjJ, \mu I}=\Gamma^{(1)}_{iIjJ,\mu I}+ \Gamma^{(1)}_{jJ iI, \mu I} \quad \text{for} \quad I>J\, .
%\end{equation}
%and 
%\begin{equation}
%    \Gamma'^{(1)}_{iIjJ, \mu J}=\Gamma^{(1)}_{iIjJ,\mu J}+ \Gamma^{(1)}_{jJ iI, \mu J} \quad \text{for} \quad I>J\, .
%\end{equation}
%Here we need less storage memory for  $\Gamma'^{(1)}_{iIjJ, \mu I}$ (as it is stored for $I>J$ in comparison to $\Gamma^{(1)}_{iIjJ, \mu I}$ (as it is %stored for all atom pairs).
 Similarly, for the second term, we have
 \begin{equation} \label{Coulomb_force}
   \big\langle \Gamma^{(2)} \dfrac{dV}{d\mathbf{R}_{A}} \big \rangle_{\langle \calI, \calJ \rangle} =  \sum_{\mu\in I} \sum_{\nu\in J} \Gamma^{(2)}_{\mu(\calI), \nu(\calJ)} \dfrac{dV_{\nu(\calJ), \mu(\calI)}}{d\textbf{R}_{A}} 
\end{equation}
Briefly, by exploiting the sparsity and symmetry properties of the intermediate quantities, the RPA force calculations can be substantially sped up, and 
the storage requirement is considerably reduced. The key operations discussed above are summarized in the algorithm presented in Algorithm.~\ref{alg:RPA_force_algorithm1}.

\AtBeginEnvironment{algorithmic}{\small}
%\begin{figure*} 
	\begin{minipage}{1.0\linewidth}
  \begin{algorithm}[H] 
			\caption{Flowchart for efficient evaluation of the  $O^{\mu}_{nm}$ tensor and \resizebox{0.075\textwidth}{!}{$\langle \Gamma^{(1)}\dfrac{dC}{dR} \rangle$}.
				Here $N_{at}$, $N_b$, $N_{occ}$, $N_{unocc}$, $N_{aux}$, $\mu (\calI)$, and $i(\calI)$ are the numbers of atoms, the AO basis functions,
				the occupied states, the unoccupied states, the total ABFs, and an ABF and AO belonging to the atom $\calI$, respectively. 
                    In practical
				calculations, $N_{aux}>N_b=N_{occ}+N_{unocc}>N_{unocc}>N_{occ}$. For calculation of $O^{\mu}_{ij}$,
                  only $N$$_{-}pairs = N_{at}(N_{at}+1)/2$
                   atomic pairs are needed, under the restriction of $\calI \geq \calJ$. }  \label{alg:RPA_force_algorithm1}
			\begin{algorithmic}[1]
				%\vspace{4pt}
				%\Statex \hspace{8.5cm}	%\textbf{Scaling Behavior}
				%\vspace{8pt}
                \State $O^{\mu(\calU)}_{nm}=0$; \quad \quad  $F^{\text{RPA}}_{c,A}=0$
				\vspace{4pt}
				\For {$\calK \gets 1$ to $N$$_{-}pairs$}
				\vspace{4pt}
                \State  $\cal{I}$: first atom in pair, $\cal{J}:$ second atom in pair, $\cal{I}\geq \cal{J}$
                \vspace{4pt}
                \State Compute $C^{\nu (\cal{I})}_{i(\calI), j (\cal{J})}$ \quad \& \quad $C^{\nu (\cal{J})}_{i(\calI), j (\cal{J})}$
                \vspace{4pt} 
                \State   $S^{\nu (\calI)}_{i(\calI),m} \gets \sum_{j,\calJ}C^{\nu (\calI)}_{i(\calI),j(\calJ)}c^{m}_{j(\calJ)}$   ~~~~~(cf. Eq.~\ref{Eq:C_times_KSvec})
                \vspace{4pt}
                \State   $S^{\nu (\calI)}_{j(\calJ),m} \gets \sum_{i,\calI}C^{\nu (\calI)}_{i(\calI),j(\calJ)}c^{m}_{i(\calI)}$ ~~~~~~(cf. Eq.~\ref{Eq:C_times_KSvec})
                \vspace{4pt}
                \State   $S^{\nu (\calJ)}_{i(\calI),m} \gets \sum_{j,\calJ}C^{\nu (\calJ)}_{i(\calI),j(\calJ)}c^{m}_{j(\calJ)}$
                \vspace{4pt}
                \State   $S^{\nu (\calJ)}_{j(\calJ),m} \gets \sum_{i,\calI}C^{\nu (\calJ)}_{i(\calI),j(\calJ)}c^{m}_{i(\calI)}$
                \vspace{4pt}	
				\EndFor
				\vspace{4pt}
               % \State  $O^{\nu(\calV)}_{nm} \gets \sum_{i,\calI}c_{i(\calI),n}O^{\nu(\calV)}_{i(\calI), m}$
                \vspace{4pt}
                \State  $O^{\mu(\calU)}_{nm} \gets \sum_{j,\calJ} \sum_{\nu,\calV} c^{n}_{j(\calJ)} ~ S^{\nu(\calV)}_{j(\calJ), m} %O^{\nu (\calV)}_{nm} 
                ~\Big[V^{\frac{1}{2}}\Big]_{\nu(\calV) \mu(\calU)}$   ~~~(cf. Eq.~\ref{eq:O_integrals_1})
                %\vspace{2pt}
                %\State $F^{\text{RPA}}_{c,A}=0$
                \vspace{6pt} 
                %\resizebox{0.5\textwidth}{!}{}
                \State $\overline{\Gamma}^{(1)}_{i(\calI),m,\mu (\calU)}=\sum_{\nu, \calV} \sum_{n}^{unocc}  V^{\frac{1}{2}}_{\mu (\calU), \nu(\calV)} \Upsilon^{\nu(\calV)}_{mn}c^{n}_{i(\calI)}$ ~~~(cf. Eq.~\ref{eq.gamma1})
                \vspace{12pt}
                \For {$\calK \gets 1$ to $N_{-}pairs$}
				\vspace{4pt}
                \State  $\calI$: first atom in pair, $\calJ:$ second atom in pair, $\calI>\calJ$
                \vspace{4pt}
                \State Compute $\dfrac{dC^{\mu (\calI)}_{i(\calI), j(\calJ)}}{dR_{\calI\calJ}}$ \quad \& \quad $\dfrac{dC^{\mu (\calJ)}_{i(\calI), j(\calJ)}}{dR_{\calI\calJ}}$
                \vspace{4pt} 
                \State $\Gamma^{(1)}_{i(\calI),j(\calJ), \mu (\calI)} \gets \sum_{m}^{occ}c^{m}_{i(\calI)} \overline{\Gamma}^{(1)}_{j(\calJ),m, \mu (\calI)}$  ~~~(cf. Eq.~\ref{eq.gamma1})
				\vspace{4pt}
                \State $\Gamma^{(1)}_{i(\calI),j(\calJ), \mu (\calI)} \gets \Gamma^{(1)}_{i(\calI),j(\calJ), \mu (\calI)}+ \sum_{m}^{occ} \overline{\Gamma}^{(1)}_{i(\calI),m, \mu (\calI)} c^{m}_{j(\calJ)}$    ~~~(cf. Eq.~\ref{eq.gamma1})
                \vspace{4pt}
                \State $F^{\text{RPA}}_{c,\calI}  \gets $ $F^{\text{RPA}}_{c,\calI} -2\sum_{i,\calI,~j,\calJ,~ \mu,\calI}\dfrac{dC^{\mu (\calI)}_{i(\calI), j(\calJ)}}{dR_{\calI \calJ}} \Gamma^{(1)}_{i(\calI),j(\calJ), \mu (\calI)} $
                %\vspace{2pt}
                \State $F^{\text{RPA}}_{c,\calJ}  \gets $ $F^{\text{RPA}}_{c,\calJ} +2\sum_{i,\calI,~ j,\calJ,~ \mu,\calI}\dfrac{dC^{\mu (\calI)}_{i(\calI), j(\calJ)}}{dR_{\calI \calJ}} \Gamma^{(1)}_{i(\calI),j(\calJ), \mu (\calI)} $
               \vspace{2pt} 
                \State $\Gamma^{(1)}_{i(\calI),j(\calJ), \mu (\calJ)} \gets \sum_{m}^{occ} c^{m}_{i(\calI)} \overline{\Gamma}^{(1)}_{j(\calJ),m, \mu (\calJ)}$  ~~~(cf. Eq.~\ref{eq.gamma1})
				\vspace{4pt}
                \State $\Gamma^{(1)}_{i(\calI),j(\calJ), \mu (\calJ)} \gets  \Gamma^{(1)}_{i(\calI),j(\calJ), \mu (\calJ)} +\sum_{m}^{occ} \overline{\Gamma}^{(1)}_{i(\calI),m, \mu (\calJ)} c^{m}_{j(\calJ)}$    ~~~(cf. Eq.~\ref{eq.gamma1})
                \State $F^{\text{RPA}}_{c,\calI} \gets F^{\text{RPA}}_{c,\calI} -2\sum_{i,\calI,~ j,\calJ,~ \mu,\calJ}  \dfrac{dC^{\mu (\calJ)}_{i(\calI), j(\calJ)}}{dR_{\calI \calJ}} \Gamma^{(1)}_{i(\calI),j(\calJ), \mu (\calJ)} $
                \vspace{3pt}
                \State $F^{\text{RPA}}_{c,\calJ} \gets F^{\text{RPA}}_{c,\calJ} +2\sum_{i,\calI,~ j,\calJ,~ \mu,\calJ}  \dfrac{dC^{\mu (\calJ)}_{i(\calI), j(\calJ)}}{dR_{\calI \calJ}} \Gamma^{(1)}_{i(\calI),j(\calJ), \mu (\calJ)} $
                %\vspace{8pt}	
				\EndFor
			\end{algorithmic}
		\end{algorithm}
	\end{minipage}
%\end{figure*}
\subsection{Computational details}
The improved algorithm for RPA gradient calculations as discussed above has been implemented in the FHI-aims code package \cite{Blum/etal:2009,Havu/etal:2009,Ren/etal:2012}.  In our calculations, we consider water clusters of size $n=$21,~22,~25 for determining the energy ordering of low-lying isomers. A modified Gauss-Legendre frequency grid with sufficient mesh points is used for frequency integration in RPA force and single-point energy calculations. %\XR{Is this sufficient?}, while for single-point energies, we have used 16 grid points for frequency integration %(\XR{Would be good to explain why making such choices.})  %For the RPA, RPA+rSE and MP2 calculations performed  with the frozen-core approximation unless. 
Frozen-core approximation is used for RPA, RPA+rSE and MP2 calculations unless otherwise stated. The convergence criterion for geometry relaxation is set to  $10^{-2}$ eV/\AA. In FHI-aims both NAO and GTO basis sets can be used. In the present work, to facilitate the comparison with quantum chemistry literature, the GTO basis sets cc-pVTZ short (TZ), aug-cc-pVTZ (aTZ), cc-pVQZ (QZ)  are used in the calculations. Furthermore, for RPA single-point energy calculations, an additional $5g$ hydrogen-like functions (with effective charge Z = 6) is used to generate extra ABFs (the “for\_aux” tag in FHI-aims)\cite{Ihrig/etal:2015,Ren/etal:2021} to improve the accuracy of local RI. The gas-phase equilibrium geometry  of H$_{2}$O monomer with  TTM2-F has ($R_{O-H}$) 0.9578 \AA, and  ($\theta$) 104.51$^{\circ}$ \cite{Xantheas_2006}. 
%\XR{Why specially mention
%the TTM2-F water molecule geometry? How about other methods?} \textcolor{red}{ This is because for other methods we can %get a relaxed structure by using FHI-aims. This is needed for Table 3 last two line data. As in literature there are %many values those are used. This is just for remembrance to reproduce our own results}

%\subsection{Accuracy check}
%\section{Implementation Details}

%\subsection{Scaling behavior of the computational cost}
\section{Results and Discussions}

The inter-molecular interaction between water molecules is of great interest in many fields of science. As mentioned above, the RPA is an appealing approach for describing the energetic and structural properties of water clusters. Developing efficient algorithms for RPA gradient calculations is hence a key step towards turning the RPA into a powerful tool for simulating properties of water systems. In this section, we present the major results of the
present work, including a benchmark comparison with literature results, the scaling behavior of our RPA gradient implementation with respect to
system size, and the energy ordering and structural parameters of a sequence of water clusters. 
%Herein, we discuss the scaling behavior, energy ordering of water clusters, the structure parameters and the number of H--bonds in  water clusters.

\subsection{Benchmark results for the (H\texorpdfstring{$_{2}$}{TEXT}O)\texorpdfstring{$_{20}$}{TEXT} clusters}
\label{sec:water20_comparison}
To start with, we examine the numerical precision of our RPA calculations for medium-sized water clusters by comparing our RPA results for 
the low-lying isomers of  $(\text{H}_{2}\text{O})_{20}$ clusters with those of Chedid \textit{et al.} \cite{Chedid/Jocelyn/Eshuis:2021}, obtained using Gaussian cc-pVTZ and cc-pVQZ basis sets. In Table~\ref{low_energy_isomers_of_water20}, 
we present the RPA@PBE total energies for four isomers of  $(\text{H}_{2}\text{O})_{20}$ clusters calculated in the present work,
in comparison with those reported in Ref.~\citenum{Chedid/Jocelyn/Eshuis:2021}. 
In our calculations, two sets of geometries are used, i.e., the geometries provided along with the WATER27 testset, obtained at the hybrid functional B3LYP level, and the geometries fully relaxed at the level of RPA@PBE. The former set of results are in excellent agreement with those reported
in Ref.~\citenum{Chedid/Jocelyn/Eshuis:2021}. The total energies can differ up to 1-2 mH, but the energy differences between the different
isomers are significantly below 0.1 mH, indicating that the original WATER27 geometries were also used in the calculations of Ref.~\citenum{Chedid/Jocelyn/Eshuis:2021}.
%It can be seen that the differences of the RPA@PBE total energies obtained using our implementation in FHI-aims 
%differ from those in Ref.~~\citenum{Chedid/Jocelyn/Eshuis:2021} by $\sim 1$ mH or less for all the isomers and two basis sets. 
Naturally, the energy orderings among the four isomers 
predicted by the two implementations are consistent with each other, ending up with a sequence of ``Edge-sharing'' $<$ ``Face-sharing'' $<$ ``Face-cubes'' $<$ ``Dodecahedron'' \cite{Mezei_2016,Chedid/Jocelyn/Eshuis:2021}, if the cc-pVQZ basis set is used. However, as shown in Table~\ref{low_energy_isomers_of_water20}, the energy ordering of the ``Face-sharing''
and ``Face-cubes'' isomers will be swapped when using the smaller cc-pVTZ basis set. Again, both implementations
consistently predicted such a basis set dependence of the RPA energy ordering, which 
signifies that high-quality basis sets are needed to obtain even qualitatively reliable RPA
results. For comparison, in Table~\ref{low_energy_isomers_of_water20} we also present the RPA@PBE results obtained using the RPA geometries. 
Now the energy differences (shown in parenthesis) obtained using the WATER27 and RPA geometries differ by 0.3 - 2 mH, but the energy ordering does not
change. Finally, as a side remark, one may notice that the RPA energies obtained using RPA geometries are slightly higher than RPA energies with WATER27 geometries, in contrast with one would expect.  This is because the RPA geometries are relaxed in the all-electron manner, while 
the RPA energies are obtained using the frozen-core approximation.

%it can be seen that energy ordering obtained by using ``FHI-aims" is  consistent with that provided by Chedid \textit{et al.} \cite{Chedid/Jocelyn/Eshuis:2021} for both kinds of basis set.  Further, with the cc-pVTZ basis set the single point energies are higher than Chedid \textit{et al.} values and with cc-pVQZ is smaller than Chedid \textit{et al.} values. The correct energy ordering of $(\text{H}_{2}\text{O})_{20}$ isomer is Edge-sharing $<$ Face-sharing $<$ Face-cubes $<$ Dodecahedron\cite{Mezei_2016,Chedid/Jocelyn/Eshuis:2021}, the cc-pVTZ basis set don't predict the correct energy ordering for $(\text{H}_{2}\text{O})_{20}$, rather cc-pVQZ predict the correct energy ordering for $(\text{H}_{2}\text{O})_{20}$ cluster. From above discussion, we conclude that cc-pVQZ basis can be trusted for the prediction of correct energy ordering of bigger cluster than $(\text{H}_{2}\text{O})_{20}$.  \\

\begin{table*} [ht!] 
    	\caption{ \label{low_energy_isomers_of_water20} Comparison of the RPA@PBE energies (in Hartree) for the four $(\text{H}_{2}\text{O})_{20}$ isomers with the literature values reported by Chedid \textit{et al.} \cite{Chedid/Jocelyn/Eshuis:2021}, using both cc-pVTZ and cc-pVQZ basis sets. The energy differences from the lowest-energy (Edge-sharing) isomer are given in parenthesis. In the present work,
     both the original WATER27 geometries (indicated by the superscript $b$) and the fully-relaxed all-electron RPA geometries (indicated by $c$) are used.}
   %  In our calculations, the geometry relaxations were done using the LRI approximation, while the single-point energy calculations were done using
    % both LRI and RI-V approximations.}
     % and frozen-core approximation.For the structure relaxation the frozen-core approximation is not used. \XR{why?}}
      \scalebox{.83}{
    	\begin{tabular}{lllll}
    		\hline\hline 	
    	\multicolumn{1}{c}{\multirow{2}{*}{basis set}} &
    		\multicolumn{1}{c}{\multirow{2}{*}{Edge-sharing}} & \multicolumn{1}{c}{\multirow{2}{*}{Face-sharing}} &    \multicolumn{1}{c}{\multirow{2}{*}{Face-cubes}} &    
      \multicolumn{1}{c}{\multirow{2}{*}{Dodecahedron}} \\\\
      \hline
       cc-pVTZ$^a$ & $-$\bf{1529.399108} & $-$1529.397262 (0.001846) &  $-$1529.397983 (0.001125)   &    $-$1529.378167 (0.020941) \\
        cc-pVQZ$^a$ & $-$\bf{1530.258903} & $-$1530.256714 (0.002189) &  $-$1530.256446 (0.002457)   &    $-$1530.241595 (0.017308) \\ [1ex]

       cc-pVTZ$^b$ & $-$\bf{1529.401778} & $-$1529.399957 (0.001821)  & $-$1529.400688 (0.001090) & $-$1529.380768 (0.021010)\\
       cc-pVQZ$^b$ &  $-$\bf{1530.260364} & $-$1530.258183 (0.002181) & $-$1530.257954 (0.002410) & $-$1530.243048 (0.017316)\\[1ex]
       
 %      cc-pVTZ$^b$ & $-$\bf{1529.401513} & $-$1529.399258 (0.002255) & $-$1529.399692(0.001821) & $-$1529.381262(0.020251)   \\ 
       cc-pVTZ$^c$ &  $-$\bf{1529.401790}  &  $-$1529.399549 (0.002241) &  $-$1529.399979 (0.001811)  &   $-$1529.381531 (0.020259) \\
       cc-pVQZ$^c$& $-$\bf{1530.258408}  & $-$1530.255937 (0.002471) & $-$1530.255528 (0.002880) & $-$1530.243154 (0.015254)\\ 
       %cc-pV5Z$^a$ & $-$1530.592364 & $-$1530.589932 &  $-$1530.589432   &    $-$1530.576983 \\ 
%       \textcolor{red}{cc-pVQZ} &     \textcolor{red}{$-$\bf1530.258160}        &    \textcolor{red}{$-$1530.255733} (0.002427)                        & \textcolor{red}{$-$1530.255204} (0.002956) &     \textcolor{red}{$-$1530.242845} (0.015315) \\
    		\hline\hline 
      $^a$Ref~\citenum{Chedid/Jocelyn/Eshuis:2021}& \multicolumn{2}{l}{~~~$^b$This work (WATER27 geometries) )} & $^c$This work (RPA geometries) \\
  \end{tabular}}
    \end{table*}

\subsection{Scaling behavior of the computational cost}
\begin{figure}[ht!] 
	\centering
	\begin{minipage}{0.8\textwidth}
		\includegraphics[width=1.0\textwidth]{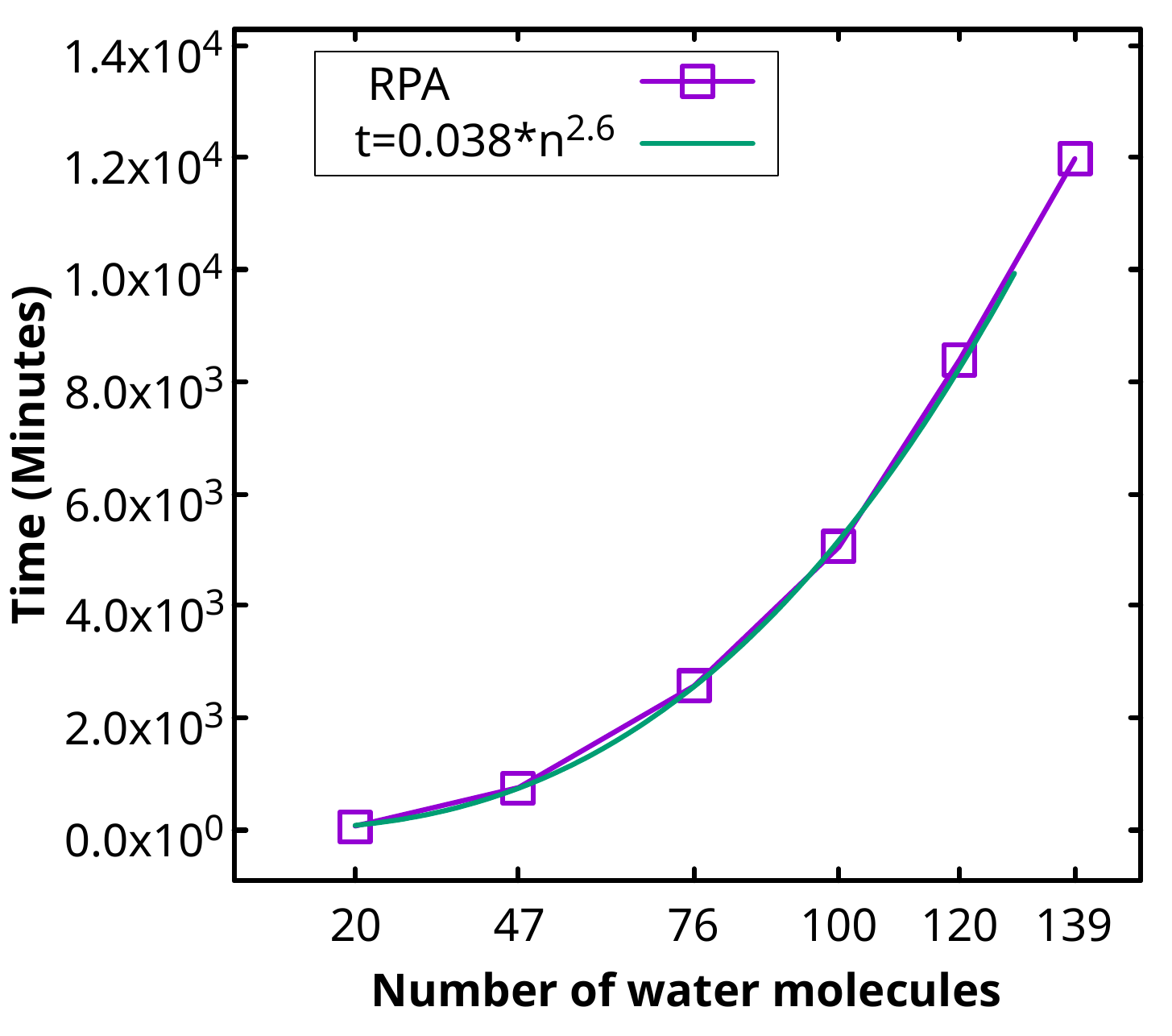} % first figure itself
	\end{minipage}
	\caption{ Wall-clock timings (in minutes) of one RPA relaxation step as a function of the water cluster size $n$. 
          The Gaussian cc-pVTZ 
           basis sets are used. The calculations were performed on 320 CPU cores. 
    A polynomial fit of the scaling behavior of the computational times is also added to the graph.
	} 
	\label{fig:time_scaling}
\end{figure}

The computational efficiency of our RPA gradient implementation has been significantly improved since its first publication 
in Ref.~\citenum{Muhammad_2022}.  
To check the scaling behavior of our improved implementation, we have considered water clusters (H$_{2}$O)$_{n}$
of increasing size ranging from $n=20$ to 139.
%(H$_{2}$O)$_{76}$,  (H$_{2}$O)$_{100}$, (H$_{2}$)O$_{120}$ and (H$_{2}$O)$_{139}$ by using Gaussian cc-pVTZ basis sets. 
%\XR{Can we expand the range of the cluster size, say from $n=20$?}
In Fig.~\ref{fig:time_scaling}, we plot the computational timings for one iteration of RPA geometry relaxation as a function
of $n$. A polynomial fit of the data points shows that the scaling behavior of the computational time is well described by
$t(n)=bn^{\alpha}$
%We perform one relaxation step calculation for the scaling test with PBE potential. 
%To judge the scaling behavior we have used the power-law fitting given as 
%\begin{equation}
%	t(n)=bn^{\alpha}
%\end{equation} 
with $\alpha=2.6$. Such a sub-cubic scaling is a significant improvement over the original $O(N^4)$-scaling algorithm, achieved by
further exploiting the sparsity of the integrals offered by the atomic-orbital basis sets and LRI scheme, as described in
Sec.~\ref{Sec:RPA_ene_LRI} and \ref{Sec:force_with_LRI}.

%Here, the scaling behavior is provided by using four data points for H$_{2}$O$_{76}$,  H$_{2}$O$_{100}$, H$_{2}$O$_{120}$ and H$_{2}$O$_{139}$. In %Figure.~\ref{fig:time_scaling}, we can see the scaling plot which shows that the overall scaling has $2.6$ exponent but the DFPT part is relatively %computationally expansive. 

\subsection{Basis set convergence}
\begin{figure}[ht!] 
	\centering
	\begin{minipage}{0.6\textwidth}
		\includegraphics[width=1.0\textwidth]{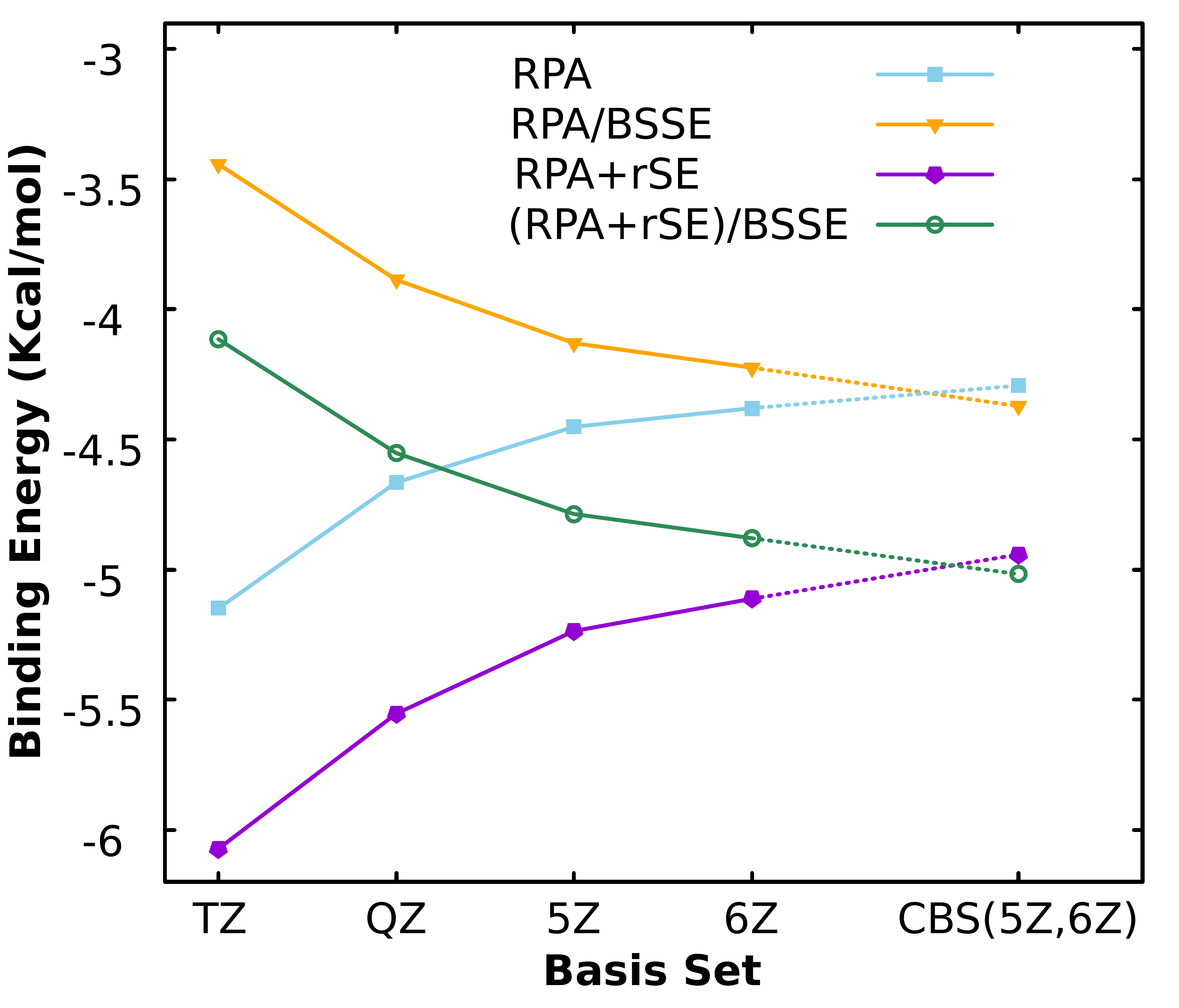} % first figure itself
	\end{minipage}
	\caption{ Convergence of the RPA@PBE and (RPA+rSE)@PBE binding energies for the water dimer 
 with respect to the basis size. The Dunning's cc-pV$n$Z basis sets are used in the tests. Binding energies with and without counterpoise
 corrections are both shown.} 
	\label{fig:water_dimer_basis_convergence}
\end{figure}

The basis set convergence of the RPA for water clusters has been thoroughly examined in Ref.~\citenum{Chedid/Jocelyn/Eshuis:2021} based on the
benchmark results on the
WATER27 \cite{bryantsev_2009} testset, which contains neutral, protonated, and deprotonated water clusters up to $n=20$.
There, it is found that the RPA underbinds the water clusters substantially at finite basis sets, and this underbinding behavior 
becomes even more severe as
one approaches the complete basis set (CBS) limit. The RPA binding strength for water clusters, 
if not corrected for the basis set superposition errors (BSSEs), becomes gradually weaker as the basis size increases.  
On the contrary, if the BSSEs are corrected, the RPA
binding energies converge from the opposite direction. Such behaviors for the binding energies are illustrated in Fig.~\ref{fig:water_dimer_basis_convergence} for the water dimer case.
The final extrapolated CBS(5,6) results with and without counterpoise corrections are fairly close, differing only 
by approximately 0.1 kcal/mol for the water dimer (cf.  Fig.~\ref{fig:water_dimer_basis_convergence}). 

Our interest here is to investigate what happens
for the RPA+rSE method which has been shown to largely remedy the underbinding problem of the RPA for weakly bonded molecules \cite{Ren/etal:2013}.
In Table~\ref{mean_deviation_WATER27}, we present the deviations of both the RPA and RPA+rSE (BSSE-uncorrected) binding energies,
obtained using both def2-QZVPP and cc-pV5Z/cc-pV6Z basis sets, from 
the reference values for WATER27 testset. The reference values are obtained using the coupled cluster theory with single, double, and perturbative
triple excitations (CCSD(T)), with dedicated efforts to achieve convergence with respect to basis size.
We note that, in this work, the binding energies of water clusters are defined as
 \begin{equation}
     E_\text{b} = -\left[E((\text{H}_2\text{O})_n) - nE(\text{H}_2\text{O})\right]
 \end{equation}
 and hence a larger positive number means stronger binding and lower total energy of an isomer. As shown in Table~\ref{mean_deviation_WATER27}, our RPA results are in excellent agreement with those reported in Ref.~\cite{Chedid/Jocelyn/Eshuis:2021}. The RPA+rSE results, first calculated 
 in the present work, show interesting features. At finite basis sets, the RPA+rSE, without correcting BSSEs, exhibits opposite behavior as RPA, i.e., it overbinds the water clusters. However, the amount of overbinding gets smaller as the basis size increases, and the extrapolated CBS(5Z,6Z) results on average only deviate by about 1.4 kcal/mol from the reference values. Such an accuracy is on a paar with the performance of
 the state-of-the-art double hybrid functional DSD-BLYP \cite{Kozuch/etal:2010} with additional dispersion corrections. Unfortunately, it should be noted that the performance of RPA+rSE, similar to RPA, has a strong basis set
 dependence, and such an accuracy can only be achieved at the CBS limit.
\begin{table*} [ht!] 
    	\caption{ \label{mean_deviation_WATER27} Mean deviations (MDs), mean absolute deviations (MADs), and maximum
absolute deviations (Max) in kcal/mol of the reference energies provides by Manna \textit{et al}.\cite{manna_2017}  for the WATER27
benchmark set structures \cite{bryantsev_2009} with (optimized geometries B3LYP/6-311++G(2d,2p) for various methods using def2-QZVPP(QZVPP) basis set. The negative error means the binding energies are underestimated.}
      \scalebox{.99}{
    	\begin{tabular}{lrrr}
    		\hline\hline 	
    	\multicolumn{1}{c}{\multirow{2}{*}{Method}} &
    		\multicolumn{1}{c}{\multirow{2}{*}{MD}} & \multicolumn{1}{c}{\multirow{2}{*}{MAD}} &  \multicolumn{1}{c}{\multirow{2}{*}{MAX}}  \\\\
      \hline
      RPA@PBE [QZVPP]$^a$ & $-$2.90 &  4.90  & 16.400 \\
       RPA@PBE [QZVPP]$^b$ & $-$2.74 &  5.02  & 16.339 \\
       (RPA+rSE)@PBE [QZVPP]$^b$ & 7.89 & 7.89 & 18.190\\
        RPA@PBE [5Z]$^b$ & -3.78 & 5.31 & 18.766 \\
        RPA@PBE [6Z]$^b$ & -5.52 &  5.74 & 21.540\\
        (RPA+rSE)@PBE [5Z]$^b$ &  6.76 &  6.76 & 15.351\\
        (RPA+rSE)@PBE [6Z]$^b$ &  4.40 &  4.40 & 10.864\\
       RPA@PBE [CBS(5Z,6Z)]$^a$ & -7.40 & 7.40 & 25.900 \\
       RPA@PBE [CBS(5Z,6Z)]$^b$ & -7.64 & 7.64 & 25.820 \\
       (RPA+rSE)@PBE [CBS(5Z,6Z)]$^b$ & 1.40  & 1.41  & 4.221\\
       M06-2X-D3(0)       &  3.40  &  3.70   &  10.000\\
       $\omega$B97X-D3(0) &  2.20 & 2.30  & 7.400\\
       DSD-BLYP-D3(0)     &  1.20 & 1.30  & 5.500 \\
    		\hline\hline 
      \multicolumn{4}{l}{$^a$Ref~\citenum{Chedid/Jocelyn/Eshuis:2021} ~~~$^b$This work (results obtained using FHI-aims)}\\
    	\end{tabular}
      }
    \end{table*}

\subsection{ Energy hierarchy for low-lying isomers of (H\texorpdfstring{$_{2}$}{TEXT}O)\texorpdfstring{$_{n}$}{TEXT} clusters with  \texorpdfstring{$n$}{TEXT}=21, 22, 25}
The energy landscapes of putative and low-lying minima of water clusters of $n=3$-$25$ 
have been thoroughly studied in Ref.~\citenum{Rakshit_2019} using
the TTM2.1-F force field (Thole Type Model version 2.1 for Flexible monomers) \cite{Burnham_2002}. Our concern here is if and how the energy ordering of the low-lying isomers determined by the force field will change when the RPA method is applied. 
As an illustration, here we only consider a collection of 9 isomers of water clusters of size  $n=$21, 22, and 25, taken from  \href{https://sites.uw.edu/wdbase/database-of-water-clusters}{the Database of Water Cluster Minima} \cite{Rakshit_2019}. 
%In our dataset, we only considered the cluster structures within $1$  KJ/mol of the force field, Thole Type Model version 2.1 for Flexible monomers (TTM2.1-F)\cite{Burnham_2002}  putative minima.
These include two low-lying isomers of $n=21$ and 22, and five low-lying ones of $n=25$.
 For these isomers, we have fully relaxed the structures using PBE, PBE0, and RPA with cc-pVQZ basis set. The (RPA+rSE)@PBE energy calculations are performed using the relaxed RPA@PBE structures. The energy calculations for these methods are performed 
 using the same cc-pVQZ basis set.  The obtained PBE, PBE0, RPA@PBE, and (RPA+rSE)@PBE results are presented in Table~\ref{low_energy_isomers_1} and \ref{low_energy_isomers_2}, 
 together with those of
 TTM2.1-F and MP2 methods reported in the literature \cite{Rakshit_2019}. Here, the MP2 results, taken from
 Ref.~\cite{Rakshit_2019}, are obtained using the aug-cc-pVTZ basis set.
 
 Table~\ref{low_energy_isomers_1} presents the computed binding energies for water clusters $n=21$ and 22, 
 with two isomers (denoted as $a$ and $b$) for each size. These two isomers are chosen since they are energetically the lowest ones 
 within 1 KJ/mol as determined by the TTM2.1-F force field. 
 In Table~\ref{low_energy_isomers_1}, the lower-energy isomer of the two clusters is highlighted in bold for each method and the energy difference 
 between the two isomers is given in parenthesis. 
 For the cluster size $n=21$, one can see that TTM2.1-F, PBE, and PBE0 predict that $a$ %\XR{Is there is a name for this isomer?} \textcolor{red}{No name in literature. a, b name are given by me.}
 is lowest-energy isomer, whereas MP2, RPA, and RPA+rSE instead favors the isomer $b$. This highlights 
 the importance of including non-local electron correlations in the calculations for water clusters. 
 To check the influence of the underlying geometries on the energy ordering, we carried out a crosscheck by performing RPA energy calculations 
 on top of the TTM2.1-F force-field geometry, and by
 performing MP2 calculations on top of the TTM2.1-F and RPA geometries. Interestingly, as indicated in Table~\ref{low_energy_isomers_1}, 
 if RPA or MP2 calculations were not performed on their own geometries, the isomer $a$ is again
 favored. This signifies the high sensitivity of the water cluster
 structures to the employed computational methods, which in turn influences the energy hierarchy for the
 energetically very close isomers (within 1 KJ/mol). 
 
 For $n=22$, all methods consistently predict that the isomer $a$ is the putative minimum, although the actual energy
 differences scatter a lot, depending on the computational methods and/or the employed geometries.
 Quantitatively, however, it seems that the energy differences based on RPA@PBE geometries
 ($\sim$ 1.1 - 1.5 kcal/mol) are noticeably larger than other cases (0.07 - 0.23 kcal/mol). 
 Considering the magnitude of binding energies themselves, the results yielded by different methods vary a lot.
 This is partly because the binding energies obtained using cc-pVQZ basis set are not yet converged.
 %by correlated methods such as RPA and MP2 are not yet fully converged at the level
 %
% of cc-pVQZ or aug-cc-pVTZ. 
 %\XR{It is somewhat strange than the PBE energies are even less well converged.}
 In particular, it can be noticed that RPA generally tends to underbind the water clusters, yielding
 a relatively smaller binding energies compared to other methods, e.g., MP2. Upon including the rSE correction, the magnitude of the
 binding energies is increased
 as much as 40 kcal/mol for this size of water clusters. This is similar to the case of water hexamers, 
 where adding the rSE correction to RPA increases the binding energy by more than 1 kcal/mol per H$_2$O molecule \cite{Muhammad_2022}.

\begin{table*} [ht!] 
    	\caption{ \label{low_energy_isomers_1} Binding energies $E_{b}$ (in kcal/mol) obtained by TTM2.1-F (force field), PBE, PBE0, MP2, RPA and RPA+rSE. The results of the lower-energy isomer are highlighted in bold for each method, and the binding energy differences between the two isomers are given in parenthesis. The geometries used in the energy calculations are explicitly indicated. The cc-pVQZ (QZ) basis set is used for all these calculations, except for PBE and MP2 calculations where the aug-cc-pVTZ (aTZ) basis set is also used.
     %and frozen-core (FC) approximation is used for RPA@PBE. 
      %We used an additional 5g hydrogen-like functions (with effective charge Z = 6) to generate extra ABFs (the “for\_aux” tag in FHI-aims)\cite{Ihrig/etal:2015,Ren/etal:2021} to enhance the accuracy of RI-LVL. For single-point energy 16 frequency points grid is used.  
      %\XR{These should belong to the session of computational details and not here. But we should indicate which results are from the literature
      %and which are obtained in the present work.}
      %The Gaussian basis sets cc-pVQZ and aug-cc-pVTZ are abbreviated as "QZ" and "aTZ", respectively.
      %The gas-phase equilibrium geometry  of H$_{2}$O monomer with  TTM2-F has ($R_{O-H}$) 0.9578 \AA, and  ($\theta$) 104.51$^{\circ}$ \cite{Xantheas_2006}. 
     }
     \begin{center}
     \scalebox{1.08}{
    	\begin{tabular}{lllllll}
    		\hline\hline 	
    	\multicolumn{1}{c}{\multirow{2}{*}{method}} &
    		\multicolumn{2}{c}{\multirow{1}{*}{$n=21$}} & & \multicolumn{2}{c}{\multirow{1}{*}{$n=22$}}    \\
      \cline{2-3}  \cline{5-6} 
      & \multicolumn{1}{c}{\multirow{1}{*}{$a$}} &  \multicolumn{1}{c}{\multirow{1}{*}{$b$}}  & & \multicolumn{1}{c}{\multirow{1}{*}{$a$}} &  \multicolumn{1}{c}{\multirow{1}{*}{$b$}} 
      \\
      \hline
      \multicolumn{6}{c}{\multirow{1}{*}{Own Geometries}}\\
       TTM2.1-F$^a$ &  \bf{227.785} & 227.565 (0.220)  & & \bf{239.167} & 239.099 (0.068) \\
       MP2/aTZ$^a$ & 232.618 (0.126) & \bf{232.744} & & \bf{242.833} & 242.701 (0.132)   \\
       PBE/aTZ &  \bf{228.217}   & 228.139 (0.078) & & \bf{235.183}  & 234.976 (0.207) \\  
       PBE  & \bf{259.651}  & 259.564 (0.087) & & \bf{268.978} & 268.749 (0.229) \\
       PBE/light & \bf{251.457} & 251.415 (0.042) & & \bf{259.657} & 259.262 (0.395)\\ 
       PBE/tight & \bf{227.940} & 227.867 (0.073) & & \bf{234.940} & 234.694 (0.246) \\
       PBE0 &  \bf{239.762} & 239.652 (0.110) &  &  \bf{248.541}&  248.423 (0.118) \\ [1.0ex]
       
       \multicolumn{6}{c}{\multirow{1}{*}{Geometry optimization RPA@PBE}}\\
    RPA@PBE & 210.818 (0.058) & \bf{210.876} & & \bf{219.609} & 218.209 (1.400) \\ 
    MP2 & \bf{241.947} & 241.795 (0.152) & & \bf{252.101} & 250.991 (1.110) \\
       (RPA+rSE)@PBE & 250.938 (0.049)  & \bf{250.987} & & \bf{263.539}  & 262.208 (1.459) \\
       [1.0ex]
       
       \multicolumn{6}{c}{\multirow{1}{*}{Geometry optimization TTM2.1-F$^a$}}\\
       RPA@PBE & \bf{207.612} & 207.323 (0.289) & & \bf{216.724} & 216.578 (0.146)  \\
       MP2 &  \bf{236.485} & 236.183 (0.302) & & \bf{246.829} & 246.768 (0.061) \\
    		\hline\hline 
      $^a$Ref~\cite{Rakshit_2019} \\
    	\end{tabular}
     }
     \end{center}
    \end{table*}

In Table~\ref{low_energy_isomers_2}, we further present the binding energy results for water clusters of $n=25$, obtained using various
methods. In this case, there are five low-lying isomers, for which the energy ordering is also of great interest, 
in analogy to the water hexamer case \cite{Biswajit/etal:2008,Muhammad_2022}. Again, in Table~\ref{low_energy_isomers_2},
the lowest-energy isomer is highlighted in bold, and the energy differences between other isomers and the lowest-energy one 
are given in parenthesis. For RPA@PBE energies, the binding energies are calculated under three sets of
geometries: the TTM2.1-F force-field geometry, the PBE geometry, and RPA@PBE geometry. Although the isomer $b$ is always the lowest-energy one
for all three sets of geometries, the actual energy orderings are different. For example, using its own geometries, the RPA@PBE predicts an energy
ordering of $b<e<a<c<d$, whereas a different energy ordering of $b<a<e<c<d$ is obtained if the PBE geometry is used. The energy ordering is
again different under the TTM2.1-F geometry. The situation here is different from what happens for the water hexamer, where using PBE or RPA@PBE geometries won't lead to a qualitatively different energy ordering \cite{Muhammad_2022}. A similar effect is also observed for MP2, where using RPA@PBE or TTM2.1-F geometries for MP2 calculations also leads to a different energy ordering, as can be seen from Table~\ref{low_energy_isomers_2}.
Compared to the water hexamer and dodecamer, here we are dealing with five isomers with the same structural motif (the O skeleton) 
and having even smaller energy differences (per water molecule). Hence the structural sensitivity of the energy ordering is much more 
pronounced here.

In Fig.~\ref{low_energy_isomers_1}, we plot the variation in the binding energies of the five isomers where the isomer $b$ is taken as
the reference (whose binding energy is set to zero for all methods). Here, the calculations of all methods are performed under 
their own geometries, except for MP2 and RPA+rSE for which the calculations are done under the RPA@PBE geometries. The isomers on the $x$-axis are ordered decreasingly  ($b \rightarrow e \rightarrow a \rightarrow c \rightarrow d$) according to the RPA binding energies evaluated on the RPA geometries.  Figure~\ref{low_energy_isomers_1} indicates that these isomers are very close in energy at the level of the TTM2.1-F force-field method, but show much larger variations when one goes to first-principles methods. Furthermore, PBE and PBE0 show similar variation patterns across the five isomers, which are however quite different from
the behaviors of MP2 and RPA-based methods. Among MP2, RPA, and RPA+rSE, the former two essentially yield the same results, whereas RPA+rSE shows
noticeable differences. In particular, compared to RPA and MP2, the order of isomers $a$ and $e$ is swapped in RPA+rSE. It is still not clear if
this ordering change upon including rSE correction agrees with more accurate quantum chemistry approach such as CCST(T) or not. More investigations
along this line are needed.

%results are presented with the TTM2.1-F force field method, RPA@PBE, and PBE. The RPA@PBE and PBE results are provided with QZ basis set while  TTM2.1-F  is based on the force field method. In Table.~(\ref{low_energy_isomers_2}), we observed that TTM2.1-F ordering is $a<b<c<d<e$, on the other hand RPA@PBE predicted ordering is $b<e<a<c<d$. The (RPA+rSE)@PBE follows the ordering of RPA@PBE, but with a swap of isomer $a$ and $c$. It is well known that with QZ basis set the RPA@PBE provides highly reliable energy ordering for the smaller size water clusters\cite{Chedid/Jocelyn/Eshuis:2021,Muhammad_2022}. We believe that current energy ordering predicted for (H$_2$O)$_{25}$ with RPA@PBE can be considered the Gold standard for benchmark reference. The  RPA@PBE is disadvantageous in a sense, it underbinds the binding energies. \\
\begin{table*} [ht!] 
    	\caption{ 
      Binding energies $E_{b}$ (in kcal/mol) of the five low-lying isomers of (H$_2$O)$_25$ clusters obtained by TTM2.1-F (force field), PBE, PBE0, MP2, RPA and RPA+rSE. The binding energies of the lowest-energy isomer are highlighted in bold for each method, and the differences between other isomers and lowest-energy one are given in parenthesis. The geometries used in the energy calculations are explicitly indicated. The cc-pVQZ basis set is used for all these calculations.
 %     Comparison of PBE and RPA@PBE binding energy $E_{b}$ (kcal/mol), and the energy difference %$\Delta E$ 
 %       in the parentheses in (Kcal/mol) from the lowest energy isomer with the TTM2.1-F force field method. Note that,
 % for these calculations, the structures are optimized with a given level of theory i.e., for PBE calculation the 
 % structure relaxation is with PBE, and for RPA@PBE  the structure relaxation is done with RPA@PBE. The cc-pVQZ basis set % is used for all the calculations performed here. In the first row, the cluster size $n$ is given with clustering ordered %from the lowest energy with respect to the TTM2.1-F method.
 } 
        \label{low_energy_isomers_2}
     \begin{center}
     \scalebox{.87}{
    	\begin{tabular}{llllll}
    		\hline\hline 	
    	\multicolumn{1}{c}{\multirow{2}{*}{method}} &
    		\multicolumn{5}{c}{\multirow{1}{*}{$n=25$}}  \\
      \cline{2-6} 
      & \multicolumn{1}{c}{\multirow{1}{*}{a}}  &  \multicolumn{1}{c}{\multirow{1}{*}{b}}  & \multicolumn{1}{c}{\multirow{1}{*}{c}} &  \multicolumn{1}{c}{\multirow{1}{*}{d}} & \multicolumn{1}{c}{\multirow{1}{*}{e}} \\
      \hline
       TTM2.1-F$^a$ &  \bf{277.472} &  277.451 (0.021) &  277.421 (0.051) & 277.346 (0.126) & 277.329 (0.143) \\
       PBE0 & 290.637 (0.790) & \bf{291.427} & 289.814 (1.613) & 289.711 (1.716) & 289.981 (1.446) \\[1.0ex]
       
       \multicolumn{6}{c}{\multirow{1}{*}{Geometry optimization PBE}}\\ 
       PBE  & 314.258 (0.907)  & \bf{315.165}  & 313.310 (1.855) & 313.256 (1.909) & 313.522 (1.643)  \\ 
       %RPA@PBE/QZ/V &  250.384(0.614)  &  \bf{250.998} & 250.938(0.060) & 250.716(0.282) & 250.935(0.063) \\
       RPA@PBE &  251.480 (0.614)  &  \bf{252.094} & 252.035 (0.059) & 251.813 (0.281) & 252.031 (0.063) \\
       MP2  & 289.988 (0.687) & \bf{290.675} & 290.368 (0.307) & 290.108 (0.567) & 290.458 (0.217) \\[1.0ex]
       
       \multicolumn{6}{c}{\multirow{1}{*}{Geometry optimization RPA@PBE}}\\ 
       
 %     RPA@PBE/QZ/V & 255.234(0.767)  & \bf{256.001} & 255.083(0.918) &  254.783(1.218) & 255.315(0.686) \\ 
       RPA@PBE & 255.243 (0.765) & \bf{256.008} & 255.095 (0.913) & 254.795 (1.213) & 255.328 (0.680) \\ 
       MP2 & 292.923 (0.771) & \bf{293.694} & 292.847 (0.847) & 292.486 (1.208) & 293.112 (0.582) \\
       (RPA+rSE)@PBE & 304.317 (0.890) & \bf{305.207} & 304.513 (0.694) &  304.202 (1.005) & 304.843 (0.364) \\ [1.0ex]
       
       \multicolumn{6}{c}{\multirow{1}{*}{Geometry optimization TTM2.1-F$^a$}}\\ 
       RPA@PBE  & 252.065 (0.752) & \bf{252.817} & 252.437 (0.380) & 252.264 (0.553) & 252.513 (0.304) \\
       MP2 & 286.984 (0.772) & \bf{287.756} & 287.197 (0.559) & 287.024 (0.732) &287.308 (0.448) \\
    		\hline\hline 
      $^a$Ref~\cite{Rakshit_2019} \\
    	\end{tabular}
     }
     \end{center}
    \end{table*}

\begin{figure}[ht!] 
	\centering
	\begin{minipage}{0.8\textwidth}
		\includegraphics[width=1\textwidth]{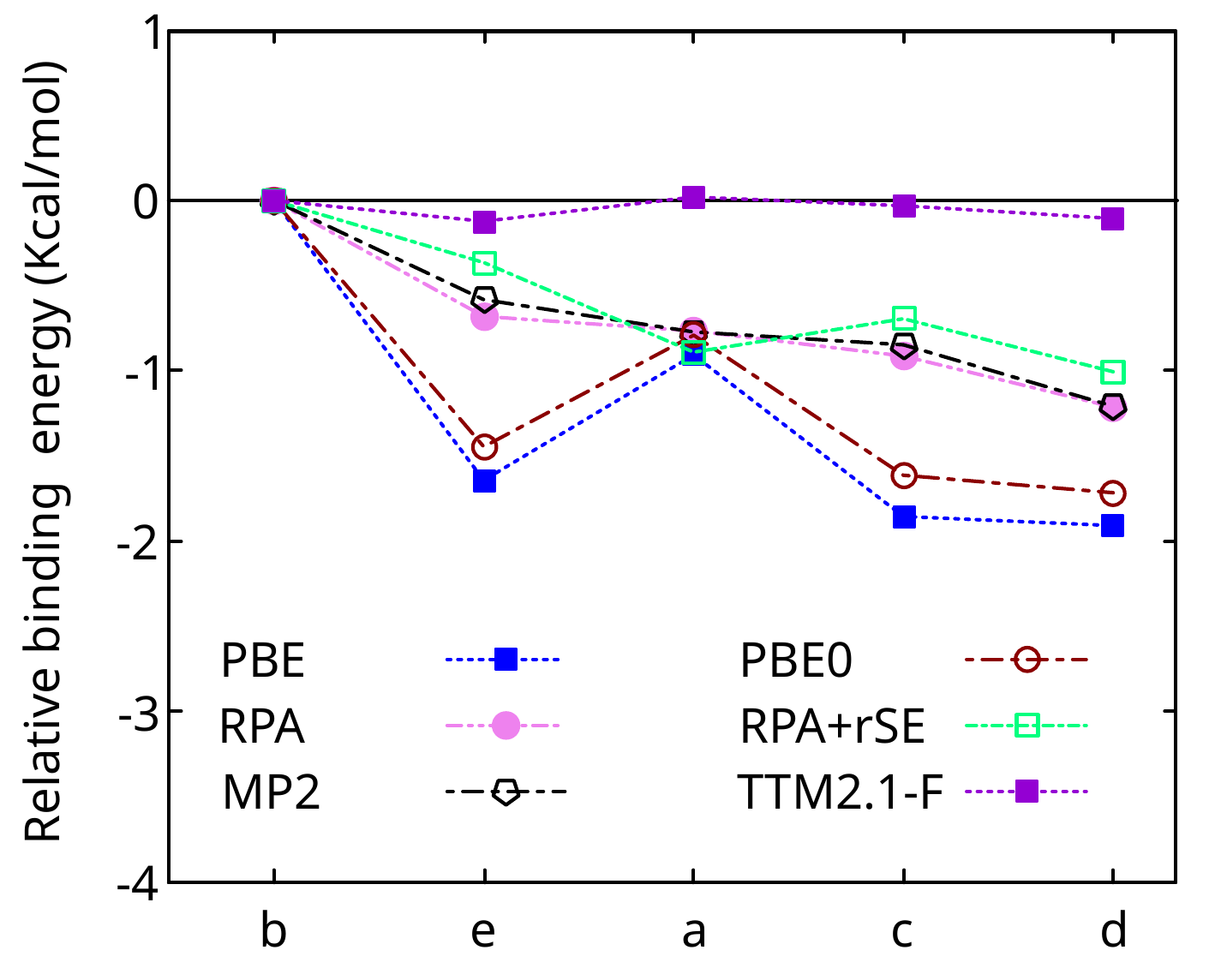} % first figure itself
	\end{minipage}
	\caption{The relative binding energies (Kcal/mol) of the isomers  of (H$_2$O)$_{25}$ with respect to isomer \textbf{b}, following the descending ordering of $b\rightarrow e \rightarrow a \rightarrow c \rightarrow d$, as given by RPA. For these calculations, the cc-pVQZ basis sets are used and the structure relaxation is performed for given methods, except for MP2 and (RPA+rSE)@PBE where RPA@PBE structures are used.} 
	\label{fig:TOC_water_25}
\end{figure}

   \section{Conclusion} 
   We present an efficient sub-cubic implementation for RPA force calculations within the framework of AO basis sets and the LRI approximation. 
   Such implementation allows us to relax the structures of water clusters containing a few tens of water molecules using RPA at the level
   quadruple-$\zeta$ basis sets. Looking into the energy hierarchy of low-lying isomers of (H$_2$O)$_{21}$, (H$_2$O)$_{22}$, and (H$_2$O)$_{25}$ clusters using RPA and other methods reveals that the energy ordering pattern predicted by RPA is quite different from those yielded by the
   force field and GGA-PBE level of theories. In particular, in contrast with the water hexamer case, the energy hierarchy given by RPA with
   force-field, PBE or RPA geometries are also different for these larger water clusters. In general, the RPA energy ordering is much closer to the MP2 one compared to the force-field or GGA-PBE results, highlighting the importance of including non-local electron correlations in water cluster
   calculations. However, the RPA itself significantly underbinds the water clusters, and such underbinding %is more severe at the CBS limit.
   becomes even more pronounced at the CBS limit.
   This underbinding is well fixed by the rSE corrections, and in fact RPA+rSE yields binding energies that is on a par with the most accurate
   double hybrid functionals, at the CBS limit. The energy ordering of the low-lying isomers can be slightly different from RPA to RPA+rSE. 
   In brief, RPA and its extensions provide a competitive class of approaches to study the structural and energetic properties of water clusters, 
   as an alternative to traditional quantum chemistry approaches as MP2. However, their strong basis-set dependence needs to be addressed to make
   them computationally economic and practical.
   
  % In conclusion, we present the scaling behavior and observe that the implementation scales with an overall $2.6$ exponent. Next, we checked the efficiency of RPA implementation by using different numbers of cores and found that code is efficient with the increase of number of cores. Further, the binding energies of low laying isomer of water clusters of size $n=21-23$ and $n=25$ is calculated and see that the QZ basis provides the correct energy ordering consistent with MP2 for $n=21$. For cluster size $n=22$, the energy ordering is accurate with TZ/QZ/NAO3Z basis set. For the cluster size $n=25$ with TTM2.1-F the ordering is a$<$b$<$c$<$d$<$e  but with RPA@PBE the ordering is b$<$e$<$a$<$c$<$d. Further, we count the H-bonds in each cluster size and see that the increase of H-bonds is not uniform on adding a single water molecule to some specific size of water cluster.    We also checked the structure parameters and concluded that $R_{O-O}$ and $R_{hb}$ underestimate for RPA@PBE in comparison to MP2. 
    
   % \begin{suppinfo}
   %  \end{suppinfo}
   %  \section*{Acknowledgments} 

    \begin{suppinfo}

%\revise{
The following file is available free of charge.
\begin{itemize}
  \item supporting\_info.pdf:
  The file contains the following items:
  \begin{enumerate}
  \item The structures of water cluster of size $n=21, 22$ and $25$, relaxed on the top of RPA@PBE using the QZ basis set and the frozen-core approximation.
  \item RPA@PBE, (RPA+rSE)@PBE, MP2, PBE, and PBE0 single point energies of the water monomer. 
  \end{enumerate}
\end{itemize}

\end{suppinfo}

\begin{acknowledgement}
  We acknowledge the funding support by the Strategic Priority Research Program of Chinese Academy of Sciences under Grant 
No. XDB0500201 and by the National Key Research and Development Program of China (Grant Nos. 2022YFA1403800 and 2023YFA1507004).  This work was also supported by the Chinese National Science Foundation Grant Nos 12134012, 12374067, 12188101, 12274254, and T2222026. The numerical calculations in this study were partly carried out on the ORISE Supercomputer.  
\end{acknowledgement}

\newpage

\section{Supporting Information for  \\ ``Efficient structural relaxation based on the random phase approximation: Applications to the water clusters
	"}   % type title between braces
  
\subsection{Structures of the  lowest-energy isomers of  water clusters (H$_2$O)$_{21}$ relaxed using RPA@PBE with the cc-pVQZ basis set. The frozen-core approximation is used.}
\begin{itemize}
\item \textbf{a}\\
\begin{equation*}
\begin{matrix}
	0.30330010     &  -3.43960684  &     -4.13185228  & O \\
    1.22225508     &  -3.13289613  &     -3.99754878  & H \\
   -0.03162000     &  -3.57525289  &     -3.23335265  & H  \\
    2.82912379     &  -5.29402016  &      1.04512867  & O  \\
    2.45009851     &  -6.17233661  &      0.87374763  & H \\
    3.65047143     &  -5.28721399  &      0.51056961  & H \\
   -0.36878557     &  -3.94711550  &     -1.37155932  & O  \\
   -1.24845807     &  -3.81536373  &     -1.00516285  & H \\
    0.25530550     &  -3.56353046  &     -0.71524376  & H \\
    1.56286253     &  -3.14213957  &      0.33484260  & O \\
    2.01256300     &  -3.98974291  &      0.62303442  & H \\
    1.38490627     &  -2.66407453  &      1.15147586  & H \\
    3.56226997     &  -8.97897038  &     -1.33714848  & O \\
    4.46317747     &  -8.58872834  &     -1.37113943  & H \\
    3.29215230     &  -9.10782219  &     -2.25990642  & H \\
    0.87752183     &  -6.34068734  &     -1.91859211  & O \\
    0.33596991     &  -5.57479678  &     -1.66077015  & H \\
    1.05242097     &  -6.85616653  &     -1.10655616  & H  \\
    5.99398644     &  -7.77819749  &     -1.29825387  & O \\
    5.74102018     &  -6.90156184  &     -0.95659820  & H \\
     
\end{matrix}
\end{equation*}

\begin{equation*}
	\begin{matrix}
6.41828236     &  -7.56099626  &     -2.14181146  & H  \\
6.85550779     &  -6.71526543  &     -3.81812614  & O  \\
6.73331315     &  -5.73403237  &     -3.75984876  & H  \\
7.68810864     &  -6.84137090  &     -4.28247120  & H  \\
4.32018865     &  -7.11110623  &     -4.67315797  & O  \\
3.94128758     &  -7.99874581  &     -4.53947481  & H  \\
5.27011470     &  -7.14815620  &     -4.45266043  & H  \\
4.68759686     &  -3.19452770  &     -5.76722906  & O  \\
4.19714486     &  -3.97539427  &     -6.14225615  & H  \\
4.98484052     &  -2.67746709  &     -6.52120699  & H  \\ 
2.94680863     &  -2.93281449  &     -3.66063871  & O  \\  
3.51322762     &  -2.78790653  &     -4.43741958  & H  \\
3.11451885     &  -3.87012874  &     -3.41148458  & H  \\
0.65952368     &  -5.40976051  &     -5.85941371  & O  \\
0.00887590     &  -5.28782862  &     -6.55807222  & H  \\
0.46429116     &  -4.69713844  &     -5.19784287  & H  \\
2.76447894     &  -9.23681254  &     -4.05188983  & O  \\
1.92637763     &  -8.72261088  &     -4.19888206  & H  \\
2.62167599     & -10.10122827  &     -4.44788997  & H  \\
5.94295352     &  -2.81647432  &     -1.30872808  & O  \\
6.59829034     &  -2.23462947  &     -0.91344809  & H  \\
5.06865437     &  -2.35455987  &     -1.21510484  & H  \\
1.82505107     &  -7.76884230  &      0.16400706  & O  \\
1.42022446     &  -8.44853284  &      0.71055365  & H  \\
2.51480417     &  -8.24389142  &     -0.38434145  & H  \\
6.43881396     &  -4.10815299  &     -3.76351101  & O  \\
6.25429635     &  -3.59828378  &     -2.95767600  & H  \\

\end{matrix}
\end{equation*}

\begin{equation*}
\begin{matrix}
5.86285168     &  -3.74669461  &     -4.45763465  & H  \\
3.27158234     &  -5.45338382  &     -2.82675717  & O  \\	
2.41337865     &  -5.80702408  &     -2.51614383  & H  \\
3.61785795     &  -6.09188329  &     -3.48548852  & H  \\
3.38563102     &  -5.32994885  &     -6.55280008  & O  \\
2.43110588     &  -5.30925583  &     -6.36538419  & H  \\
3.72725360     &  -6.03960380  &     -5.97904855  & H  \\
3.52774362     &  -1.76887844  &     -1.25466171  & O  \\
2.88497139     &  -2.19837778  &     -0.66758492  & H  \\
3.26738598     &  -2.06613279  &     -2.14868050  & H  \\
4.91324149     &  -5.30040138  &     -0.69614123  & O  \\
5.41075805     &  -4.46904281  &     -0.77519343  & H  \\
4.35495903     &  -5.33255325  &     -1.50393100  & H  \\
0.62428815     &  -7.73536134  &     -4.27795593  & O  \\
0.58013199     &  -7.27500067  &     -3.42030563  & H  \\
0.62692279     &  -7.00931027  &     -4.92531991  & H  \\
  \end{matrix}
  \end{equation*}

\newpage
\item \textbf{b} \\
\begin{equation*}
\begin{matrix}
       0.34993576  &     -6.27070293  &      1.49782220 & O\\
      -0.05873985  &     -6.50224046  &      0.64564284 & H\\
      -0.03222052  &     -5.40025962  &      1.70732970 & H\\
      -1.41606633  &     -0.20862155  &     -1.90690594 & O\\
      -0.54594755  &      0.15290533  &     -2.14738370 & H\\
      -1.35555767  &     -1.14088903  &     -2.19886890 & H\\
       1.23578574   &     0.38588841  &     -2.59906298 & O\\
       1.54904125   &     1.13619610  &     -3.11270457 & H\\
       1.57987899   &    -0.42579256  &     -3.07408045 & H\\
       2.24529422   &    -1.72115237  &     -3.70820962 & O\\
       2.99020190   &    -2.06649595  &     -3.16985309 & H\\
       1.71731409   &    -2.49470867  &     -3.96272909 & H\\
       3.32082224   &    -3.05068877  &      2.22544724 & O\\
       2.52051825   &    -2.72004714  &      2.66407755 & H\\
       3.15650905   &    -3.98851208  &      2.02667212 & H\\
       4.25853514   &    -2.68524540  &     -2.14863961 & O\\
       4.43725646   &    -2.18222046  &     -1.34008727 & H\\
       3.86664250   &    -3.50855251  &     -1.80502470 & H\\
       0.82129156   &    -2.91846892  &     -0.41318180 & O\\
       1.20214120   &    -2.02445708  &     -0.28267427 & H\\
       0.36805872   &    -3.14983757  &      0.42206024 & H\\
      -0.99842724  &     -2.84725089  &     -2.38933595 & O\\
      -0.27798275  &     -2.89721433  &     -1.72239533 & H\\
      -0.59591775  &     -3.17544386  &     -3.21327202 & H\\
      2.72416913  &     -4.72436752  &     -1.06661552 & O\\      
\end{matrix}
\end{equation*}

\begin{equation*}
	\begin{matrix}
2.98067469  &     -5.18920974  &     -0.25329331 & H\\
2.05531857  &     -4.07138282  &     -0.76727690 & H\\		
0.23731616  &      0.07214340  &      1.99675016 & O\\
-0.64192722  &      0.00583756  &      1.58846239 & H\\
0.85106783  &      0.01803250  &      1.23857213 & H\\
2.92294903  &     -5.73256277  &      1.50666608 & O\\
3.41771305  &     -6.43169700  &      1.94363211 & H\\
1.97237886  &     -6.02137586  &      1.51641801 & H\\
4.34854715  &     -1.40529678  &      0.42913437 & O\\
4.00552504  &     -2.05520400  &      1.09330512 & H\\
5.10253542  &     -0.98295738  &      0.85114508 & H\\
-2.63919908 &      -4.54475729 &      -1.02118282 & O \\
-2.14105728 &      -3.93155352 &      -1.59859530 & H\\
-2.85812939 &      -4.00689675 &      -0.24687990 & H\\
-2.20527616 &      -0.20665553 &       0.56424781 & O\\
-2.81908817 &       0.53334232 &       0.61868882 & H\\
-1.87411879 &      -0.18770532 &      -0.38218033 & H\\
-0.99019687 &      -6.62275577 &      -0.95388677 & O\\
-1.52172546 &      -7.41271820 &      -1.09475320 & H\\
-1.64094077 &      -5.87696596 &      -0.92603644 & H\\
-2.97386483 &      -2.72274767 &       1.21930679 & O\\
-3.76897946 &      -2.70835114 &       1.76075412 & H\\
-2.80186140 &      -1.78720675 &       0.97124032 & H\\
0.97670490  &     -1.98796654  &      3.43212131  &O\\
0.99781354  &     -1.77254403  &      4.36903195  &H\\
0.68513639  &     -1.16019576  &      2.96153551  &H\\
0.56704787  &     -3.91188103  &     -4.36122133  &O\\

\end{matrix}
\end{equation*}

\begin{equation*}
	\begin{matrix}
0.74622757  &     -4.74746576  &     -3.85098923  &H\\		
0.46553373  &     -4.17666897  &     -5.27978586  &H\\
1.87748663  &     -0.44423553  &     -0.11542272  &O\\
2.82945600  &     -0.60973770  &      0.02224892  &H\\
1.77454926  &     -0.01932973  &     -0.98923713  &H\\
-0.47366364 &      -3.65606243 &       1.86412181 & O\\
-1.39350400 &      -3.34874231 &       1.77881238 & H\\
-0.07102303 &      -3.13063256 &       2.58024923 & H\\
1.04795904  &     -5.99049080  &     -2.83214040  &O\\
1.71850380  &     -5.65260057  &     -2.20651579  &H\\
0.30622418  &     -6.24833925  &     -2.25967842  &H\\
\end{matrix}
\end{equation*}
\end{itemize}

\newpage
\subsection{Structures of the  lowest-energy isomers of  water clusters (H$_2$O)$_{22}$ relaxed using RPA@PBE with the cc-pVQZ basis set. The frozen-core approximation is used.}
\begin{itemize}		
\item \textbf{a} \\
\begin{equation*}
\begin{matrix}	
	-0.70068807    &    2.38062258  &     -1.42309541  & O \\
	-0.55020091    &    2.94707208  &     -0.64724220  & H \\
	-2.39495380    &    0.07236942  &      3.90276973  & H \\
	3.76140342     &   2.28346572   &     2.68853460  & O \\
	0.06636984     &   1.77600306   &    -1.41096498  & H \\
	-2.25131829    &    0.80562113  &      2.55215216 & H \\
	1.40655769     &   1.55687541   &     1.59417549  & O \\
	4.06459735     &   2.96284120   &     2.05906016  & H \\
	-0.39684418    &   -0.22068607  &      3.21774153 & H \\
	-2.86422877    &    1.02708671  &     -0.80930951 & O \\
	4.21264622     &   1.45958960   &     2.42157821  & H \\
	0.54225190     &   0.52459815   &     4.16154068  & H \\ 
	-2.62263698    &    2.25269454  &      1.64831314 & O  \\
	1.07834167     &   0.90792978   &     2.26015025  & H \\
	5.27523325     &  -0.64187751   &     1.72828962  & H \\
	1.45940017     &   0.70328096   &    -0.94341644  & O \\
	2.26206304     &   1.86927474   &     1.95405450  & H \\
	3.72229501     &  -0.65963074   &     1.67547445  & H \\
	-1.85486591    &   -1.28475681  &      0.45871043 & O \\
	-3.53030305    &    1.09020487  &     -1.50011992 & H \\
	1.95574136     &   2.66106451   &     5.32461509  & H \\
	\end{matrix}
	\end{equation*}

\begin{equation*}
	\begin{matrix}
		1.69579840     &   5.33150376   &     1.23718862  & O \\
		-2.06936550    &    1.52368000  &     -1.14485452 & H \\
		3.11660740     &   2.79960286   &     4.31140479  & H \\
		4.05551554     &   3.92180072   &     0.50946051  & O \\
		-1.84013952    &    2.79000382  &      1.43197858 & H \\	
		-2.43957812    &    3.72227077  &      4.23008532 & H \\
		0.63591022     &  -1.86069768   &    -0.25157749  & O \\
		-2.91840179    &    1.90478647  &      0.78753227 & H \\
		-3.35595120    &    2.93186019  &      3.32491934 & H \\
		-0.63380421    &    4.04419951  &      3.81213271 & O \\
		1.23445348     &  -0.24232386   &    -0.94360835  & H \\
		-2.87328380    &    1.82050102  &      5.16692197 & H \\
		-2.09863929    &   -0.06019410  &      2.98712119 & O \\
		1.44740613     &   0.93706230   &     0.01037687  & H \\
		-3.09363493    &    0.77230101  &      6.29492775 & H \\
		0.55759821     &  -0.12476428   &     3.42711493  & O \\
		-2.03509724    &   -1.07058546  &      1.38715255 & H \\
		-0.12835481    &    2.62320403  &      4.68589362 & H \\
		4.51786922     &  -0.07521177   &     1.55256886  & O \\
		-2.25563590    &   -0.55344244  &     -0.04190137 & H \\
		-0.61967306    &    1.53863180  &      5.65415499 & H \\
		2.69005880     &   3.23944856   &     5.06738651  & O \\
		1.83048200     &   5.48090358   &     2.18777174  & H \\
		4.34105643     &   0.93995579   &    -0.04002415  & H \\
	   -3.25588781     &   3.19465225   &     4.25369354  & O \\
		2.50765940     &   4.89345303   &     0.93111738  & H \\
		3.20398032     &   1.26109049   &    -1.00632609  & H \\
	\end{matrix}
\end{equation*}	

	\begin{equation*}
	\begin{matrix}
		-2.50513663  &      0.99358538  &      5.56674838 & O \\
		4.77504975   &     4.45469835   &     0.15669941  &H \\
		0.35914210   &     2.77297227   &     1.31656155  &H \\
		0.18782402   &     1.86900191   &     5.22934010  &O \\
		4.04280854   &     3.09638699   &    -0.04387112  &H  \\
		0.38675755   &     4.28867342   &     1.06795362  &H \\
		4.07725310   &     1.58444619   &    -0.71781107  &O \\
		0.58631567   &    -2.70818394   &    -0.70448673  &H \\
		1.80204358   &    -1.19787393   &     2.59245748  &H \\
		-0.24718034  &      3.53381805  &      1.13773062 & O\\
		-0.29386868  &     -1.66372881  &      0.02737669 & H\\
		1.83079331   &    -1.75594592   &     1.16697233  &H \\
		2.40161961   &    -1.59374711   &     1.93792175  &O \\
		-0.53420656  &      3.87232996  &      2.85465258 & H\\ 
		2.12750709   &     4.69836743   &     4.50026460  &H \\
		1.69934903   &     5.48756761   &     4.07847915  &O \\
		0.04326704   &     4.71905810   &     3.99729077  &H \\
		1.94570433   &     6.24353721   &     4.61974265  &H \\ 
	\end{matrix}
\end{equation*}
	
	\newpage
\item	\textbf{b} \\
	\begin{equation*}
	\begin{matrix}
	 -3.43475339   &     1.01487758   &     0.65012457 & O \\
	 -4.01884491   &     1.62825679   &     0.19466910 & H \\
	 -1.75065503   &    -6.00283767   &    -1.53895892 & H\\
	 -2.45900997   &     0.15169806   &     6.04960907 & O\\
	 -3.86208525   &     0.12285209   &     0.57800131 & H\\
	 -1.38872105   &    -5.60105189   &    -0.07440169 & H\\
	 -0.22331526   &    -0.36405295   &     2.80510715 & O\\
	 -2.64541626   &     0.49605072   &     5.16078214 & H\\
	 -0.54649594   &    -6.02972714   &     1.93807146 & H\\
	 -2.76914825   &     0.43207726   &     3.26954714 & O\\
	 -1.51970297   &    -0.09022743   &     6.00746433 & H\\
	 -1.94098043   &    -5.39481946   &     2.02295064 & H\\
	 2.49182483    &   -2.20757833    &    5.11522919  & O\\
	 -0.11511962   &    -0.61925678   &     3.74900763 & H\\
	 2.32197834    &   -2.20717453    &    2.01772269  & H\\
	 -3.80493459   &    -2.15079470   &     6.21193154 & O\\
	 -0.39975552   &    -1.20565323   &     2.32422214 & H\\
	 1.68840452    &   -0.89055510    &    2.50530732  & H\\
	 -0.03983333   &    -1.12188163   &     5.36026843 & O\\
	 -1.84493758   &     0.19270901   &     3.04330067 & H\\
	 -1.11530057   &    -2.64190409   &     0.63688088 & H\\
	 1.05858623    &   -4.64284212    &    5.07522288  & O\\
	 -3.10231224   &     0.88260800   &     2.47629450 & H\\
	 0.25680485    &   -3.05402402    &    1.32822665  & H\\
	 -4.14405038   &    -1.90790665   &     3.40242969 & O\\
		\end{matrix}
	\end{equation*}

\begin{equation*}
	\begin{matrix}
		 2.58549479    &   -1.91972612    &    4.17068712  & H\\ 
		-3.04638768   &    -3.20756517   &     3.03405102 & H\\
		-0.67049761   &     0.67355711   &     0.21501087 & O\\
		3.34693218    &   -2.05034601    &    5.52639689  & H\\
		-1.64268811   &    -3.37727264   &     2.40854192 & H\\
		1.11100487    &   -5.85105927    &    2.72884953  & O\\
		-3.35695124   &    -1.26901752   &     6.24524736 & H\\
		1.70549898    &   -4.57463173    &    1.60415043  & H\\
		-1.39704320   &    -5.27699229   &    -1.01635858 & O \\
		-4.38692362   &    -2.17667050   &     6.97741916 & H \\
		1.65507050    &   -3.83993869    &    0.20080810  & H \\
		-1.41220008   &    -6.01783982   &     1.49406708 & O \\
		-0.52359224   &    -1.97858215   &     5.38673072 & H \\
		0.61448379    &   -2.85698205    &   -1.50110169  & H \\
		2.48875503    &   -1.43218822    &    2.58442584  & O \\
		0.89248813    &   -1.36274967    &    5.49525401  & H \\
		0.19988299    &   -4.33675288    &   -1.40094051  & H \\
		-0.57557535   &    -2.54615031   &     1.44768866 & O \\
		1.62347438    &   -3.85325138    &    5.13133212  & H \\
		-0.12341862   &    -0.70275024   &    -2.44944499 & H \\
		-2.31660192   &    -3.85516541   &     2.94064074 & O \\
		0.18664717    &   -4.34842156    &    5.38479329  & H \\
		-0.38428624   &    -0.51040292   &    -0.93497555 & H \\
		1.78342966    &   -3.70846861    &    1.16595695  & O \\
		-3.64378899   &    -1.06257433   &     3.38878561 & H \\
		-2.14526883   &    -3.68112889   &    -0.87733347 & H \\
		0.97479959    &   -3.75146186    &   -1.38561734  & O \\
	\end{matrix}
\end{equation*}

\begin{equation*}
	\begin{matrix}
		-4.32969613   &    -2.05591919   &     4.34323209 & H \\
		-1.77948634   &    -2.20695777   &    -1.22619928 & H \\
		-0.31969448   &    -1.19564541   &    -1.64664258 & O \\
		-0.36307158   &     0.51100965   &     1.11862677 & H \\
		-2.24644435   &    -3.19149337   &     5.91395559 & H \\
		-2.27572201   &    -2.75594700   &    -0.59288308 & O \\
		-1.62119140   &     0.85976875   &     0.30844587 & H \\
		-1.74270616   &    -3.64155727   &     4.52919211 & H \\
		-1.43304157   &    -3.43189167   &     5.44033864 & O \\
		1.07540130    &   -5.40679738    &    3.61611632  & H \\  
		-3.73294881   &    -2.01533666   &     0.20918536 & H \\
		-4.41389514   &    -1.44501626   &     0.60668968 & O \\
		1.58244577    &   -6.67584686    &    2.88366347  & H \\
		-4.53287077   &    -1.76594450   &     1.51328853 & H \\
	\end{matrix}
\end{equation*}
\end{itemize}

\newpage
\subsection{Structures of the  lowest-energy isomers of  water clusters (H$_2$O)$_{25}$ relaxed using RPA@PBE with the cc-pVQZ basis set. The frozen-core approximation is used.}
\begin{itemize}		
	\item \textbf{a} \\
	\begin{equation*}
		\begin{matrix}
			10.40603729    &   10.50814590   &   -11.06728442 & O    \\
			9.58139714     &  10.44801855    &  -10.57546758  & H \\    
			11.08742465    &   10.79333906    &  -10.40549403 & H   \\ 
			14.75502692    &   10.48409184    &  -12.78323102 & O    \\
			14.48391170    &    9.70124746    &  -12.25338033 & H   \\ 
			15.67459675    &   10.71926659    &  -12.52603249 & H \\
		    17.14332144 &      13.25734889  &   -13.95020001 & O \\
			16.30435185 &      13.68485487  &   -14.18012014 & H \\
			17.40442686  &     12.78937782     & -14.76132496 & H  \\
			15.91800664  &      7.05328070     & -12.87002876 & O \\
			15.77809381  &      7.09203071     & -13.83043498 &  H \\
			16.68356617  &      7.63198772     & -12.72083906 & H \\
			11.00626268  &     12.06499279     & -13.26054163 & O \\
			10.64916381   &    11.62529900     & -12.46838517  & H \\
			11.90361099   &    12.37313862     & -13.00169070  & H \\
			12.34412207   &    14.99685959     & -16.22855824  &O \\
			12.08417210   &    15.90706430     & -16.39855216  &H \\
			11.53327528   &    14.54647182     & -15.87782397  &H \\
			14.95816247   &    11.89332138     & -16.62514823  &O \\
			15.92781259   &    11.81616403     & -16.67198344  &H \\
			14.65745102   &    11.08222713     & -16.15730282  &H \\

		\end{matrix}
	\end{equation*}

\begin{equation*}
	\begin{matrix}
		14.43998670   &     9.51332838      & -9.27655457  &O \\
		14.54305256   &     8.88877237      & -8.55205770  &H \\
		14.26466454   &     8.96448495     & -10.07623070  &H \\
		13.51019105    &   12.80551360     & -12.65752705  &O \\
		13.86897616    &   13.24310316     & -13.46429984  &H \\
		13.93802843   &    11.91778394     & -12.65696220  &H \\
		14.09085762   &     9.77304891     & -15.23469916  &O \\
		14.39996512   &    10.03909579     & -14.33760829  &H \\
		13.11046705   &     9.85959766     & -15.18503044  &H \\
    	18.06131826     &   8.93054391  &    -12.65668567 & O  \\
		17.80697157    &    9.84153645 &     -12.37947013 & H \\ 
		18.90525085    &   8.75647021   &   -12.22882035 & H   \\
		12.26211696    & 11.27559883     &  -9.35220192 & O   \\
		12.98818893    & 10.63755253      & -9.24276396 & H   \\
		12.70649790      & 12.11935814     &  -9.53877985 & H   \\
		10.64611337      & 11.46866026     & -17.27797446 &  O  \\
		10.13311293      & 11.21017968     & -18.04931293 & H  \\
		11.45454676      & 11.92106994     & -17.63167881 & H  \\
		14.50149402      & 13.90127112     & -14.87507013 & O \\
		13.81742618    &   14.45701003   &   -15.28779918 & H \\
		14.64043717     &  13.17227406    &  -15.51986719 & H \\
		17.67713889     &  11.66246016    &  -16.24092153 & O \\
		17.77648621    &   10.71851959    &  -15.95331882 & H \\
		18.39162795    &   11.81562065    &  -16.86668045 & H \\
		13.81961982    &   13.49264317    &  -10.12461212 & O \\
		13.72223822    &   13.39884052    &  -11.10210579 & H \\
		13.70834378    &   14.42748799    &   -9.92924557  & H \\ 
	\end{matrix}
\end{equation*}

\begin{equation*}
\begin{matrix}
	17.88598701    &    9.15285873    &  -15.44492856 & O \\
	18.03540175    &    9.05246377    &  -14.48896679 & H \\
	17.13136559    &    8.57095664    &  -15.63507921 & H \\
	16.06983340    &   11.82384876    &   -9.68725608 & O \\
	15.36941940    &   12.49393104    &   -9.74944003 & H \\
	15.59174644    &   11.00253592    &   -9.48694949  & H \\
	11.45085923    &   10.06731113    &  -15.02539989 & O \\
	11.09822896    &   10.38575297    &  -15.87475021 & H \\
	11.29232424    &   10.81280477    &  -14.40441856 & H \\
	15.54244057    &    7.59974472    &  -15.62663577 & O \\
	14.88365911    &    8.33214501    &  -15.56845587 & H \\
	15.23774179    &    7.01565108    &  -16.32697864 & H \\	
	12.78450433    &   12.83619696     & -18.04971691 & O   \\
	12.68924010   &    13.68804835    &  -17.59348849 &  H   \\
	13.61767156   &    12.47431751    &  -17.69849816 & H   \\
	11.50919851   &     8.43283713    &  -12.66473048 & O   \\
	11.04057133   &     9.07576121    &  -12.10602714 & H   \\
	11.42146059   &     8.78739987    &  -13.56255729 & H   \\
	17.15503337   &    11.34230402    &  -12.07397859 & O   \\
	16.85364021   &    11.63537240    &  -11.18353537 & H   \\
	17.18414824   &    12.12106823    &  -12.67404248 & H    \\
	10.23386876   &    13.60553492    &  -15.42244068 & O   \\
	10.16740327   &    12.89002648    &  -16.07551067 & H   \\
	10.38144838   &    13.14042492    &  -14.57960727 & H   \\
	13.94750426   &     8.28127286     & -11.56049727 & O   \\
	14.58422771   &     7.70597935    &  -12.04344566 & H   \\
	13.06756066   &     8.22014347   &   -11.99565773 & H \\
\end{matrix}
\end{equation*}

\newpage
	\item \textbf{b} \\
	\begin{equation*}
		\begin{matrix}
			   16.48813723  &      9.97857548    &  -15.02009849  &  O   \\
			17.29457224  &      9.41238436    &  -15.03162479  &  H   \\
			15.74584600  &      9.41821795    &  -15.34753980  &  H   \\
			14.81995857  &      7.33481810    &   -9.23963843  &  O   \\
			14.98276256  &      6.74452220    &   -8.49845169  &  H   \\
			15.19905444  &      8.21819828    &   -8.98103934  &  H   \\
			17.02625312  &     12.54166449    &  -15.32139621  &  O   \\
			16.95464994  &     12.84940679    &  -16.22959082  &  H   \\
			16.77785398  &     11.58735162    &  -15.33541663  &  H   \\
			12.43661267  &      8.47626339    &  -10.25611071  &  O   \\
			13.15220846  &      7.94576691    &   -9.86958896  &  H   \\
			12.69550130  &      9.39615277    &  -10.07377778  &  H   \\
			14.41950136  &      8.48234425    &  -15.71943579  &  O   \\
			14.68179853  &      7.93301128    &  -16.47916679  &  H   \\
			14.26541427  &      7.83647835    &  -14.99457939  &  H   \\
			18.65783175  &      8.45662114    &  -14.88786955  &  O   \\
			19.40582953  &      9.00501180    &  -14.58623292  &  H   \\
			18.44548613  &      7.86647427    &  -14.12886729  &  H   \\
			20.51754786  &     10.08254678    &  -13.70642456  &  O   \\
			20.09142885  &     10.97889943    &  -13.70023031  &  H   \\
			21.43068992    &   10.23271027    &  -13.96892703  &  H \\
			18.07594512    &    6.73527958    &  -12.90949342  &  O \\
			17.97270565    &    5.90253222    &  -13.40354206  &  H \\
			17.20973784    &    6.88319298    &  -12.47480825  &  H \\
			11.67892799    &    8.02962239    &  -12.74998343  &  O \\

		\end{matrix}
	\end{equation*}
	
	\begin{equation*}
		\begin{matrix}
			11.95553575    &    8.19075211    &  -11.81206091  &  H \\
			10.76623384    &    7.72980107    &  -12.69618742  &  H \\
			15.63613519    &   13.43742875    &  -13.02738759  &  O \\
			16.05907892    &   13.24810967    &  -13.88106945  &  H \\
			16.34936968    &   13.30674507    &  -12.38056427  &  H \\
			12.55544848    &   10.20806343    &  -14.36056003  &  O \\
			13.12387478    &    9.75065505    &  -14.99909076  &  H \\
			12.15178627    &    9.48584271    &  -13.84999352  &  H \\
			13.77719685    &   11.62443758    &  -12.43136491  &  O \\
			13.30016530    &   11.24653841    &  -13.20284289  &  H \\
			14.36308903    &   12.35716435    &  -12.73200118  &  H \\
			13.42313453    &   11.07461635    &   -9.87788437  &  O \\
			13.50164829    &   11.36391639    &  -10.81730553  &  H \\
			12.91173726    &   11.76084766    &   -9.43853089  &  H \\
			17.60968249    &    4.64369705    &  -14.68221349  &  O \\
			16.63883490    &    4.63824297    &  -14.88427955  &  H \\
			17.86025169    &    3.71932790    &  -14.59225138  &  H \\
			15.66662687    &    7.29130822    &  -11.83574764  &  O \\
			15.62285926    &    8.26263035    &  -11.99572213  &  H \\
			15.43552812    &    7.16804937    &  -10.89681581  &  H \\
			18.05204939    &    6.51862769    &  -16.78356630  &  O \\
			18.07191581    &    5.77962397    &  -16.15334461  &  H \\
			18.38346123    &    7.26916471    &  -16.25993530  &  H \\
			17.87397212    &   12.86606480    &  -11.33924770  &  O \\
			18.15176788    &   13.41447843    &  -10.59940646  &  H \\
			17.84944093    &   11.93896619    &  -10.99545232  &  H \\
			15.49958237    &    6.65887115    &  -17.49805712  &  O \\
		\end{matrix}
	\end{equation*}

\begin{equation*}
	\begin{matrix}
		
		16.47121433    &    6.66194886    &  -17.30177999  &  H \\
		15.42547428    &    6.55407920    &  -18.45115622  &  H \\
		15.74084827    &    9.83737064    &  -12.47564792  &  O \\
		16.05217293    &    9.91373942    &  -13.40587872  &  H \\
		14.98464713    &   10.46134869    &  -12.40899667  &  H \\
		19.32783805    &   12.43844438    &  -13.69688555  &  O \\
		18.61112339    &   12.53255521    &  -14.34656900  &  H \\
		18.91185610    &   12.66499822    &  -12.84710755  &  H \\
		17.63375347    &   10.30951785    &  -10.67624120  &  O \\
		16.97845419    &   10.12768784    &  -11.38650742  &  H \\
		18.37949413    &    9.69823028    &  -10.87094606  &  H \\
		13.87830520    &    6.70405207    &  -13.74975692  &  O \\
		14.49320953    &    6.85168972    &  -12.99886519  &  H \\
		13.01252131    &    7.00854271    &  -13.42294098  &  H \\
		19.61568964    &    8.63105576    &  -11.42408494  &  O \\
		19.23597271    &    7.83886866    &  -11.83586478  &  H \\
		20.05667064    &    9.09551115    &  -12.15421489  &  H \\
		15.69603871    &    9.76415803    &   -8.78518552  &  O \\
		16.46541002    &    9.95947751    &   -9.35420673  &  H \\
		14.98874032    &   10.34523289    &   -9.11002237  &  H \\
		15.05116506    &    4.79667362    &  -15.37544579  &  O \\
		15.06446450    &    5.30793494    &  -16.20067951  &  H \\
		14.52457582    &    5.34606185    &  -14.76783907  &  H \\
	\end{matrix}
\end{equation*}

\newpage
\item \textbf{c} \\
\begin{equation*}
	\begin{matrix}
    -7.84140253   &    -2.13833377    &    6.07704888  &  O  \\
	-6.93153680   &    -1.83820632    &    6.23802889  &  H  \\
	-8.42831064   &    -1.42271275    &    6.37063302  &  H   \\
	-10.26801852   &    -2.41923132    &    2.37942217  &  O   \\
	-9.63842094   &    -2.99514610    &    2.83964750  &  H   \\
	-10.38244148   &    -1.68724071    &    3.01245620  &  H   \\
	-8.20984207   &    -3.64779813    &    3.94180857  &  O   \\
	-8.11676787   &    -3.08214176    &    4.74956874  &  H   \\
	-8.19359367   &    -4.55379117    &    4.26418975  &  H   \\
	-5.52297367   &     0.66705417    &    0.32717203  &  O   \\
	-5.82036399   &    -0.20423908    &    0.01055106  &  H   \\
	-6.33745352   &     1.08020585    &    0.68765270  &  H   \\
	-6.39165797   &     1.08755997    &    5.66884120  &  O  \\ 
	-6.76801415   &     0.63515886    &    4.87823174  &  H   \\
	-5.88820858   &     0.39014255    &    6.12279120  &  H  \\ 
	-5.15951761   &     5.10965726    &    1.00445353  &  O  \\ 
	-5.23579392   &     5.85600524    &    0.40251079  &  H   \\
	-4.86328540   &     4.34559010    &    0.44632124  &  H   \\
	-10.13966271   &     1.58238516    &    0.49995265  &  O  \\ 
	-10.65519001   &     2.03123724    &   -0.17574917  &  H \\	
	-10.55237468   &     1.81629139    &    1.37729934  &  H  \\
	-9.72645789   &    -0.14611092    &    6.76674329  &  O   \\
	-10.29699339   &    -0.23349878    &    7.53559440  &  H   \\
	-9.42972647   &     0.79970129    &    6.74765297  &  H   \\
	-4.15485580   &    -1.50638747    &    3.96145735  &  O   \\
	\end{matrix}
\end{equation*}

\begin{equation*}
	\begin{matrix}
		      -3.92535455   &    -0.68140061    &    3.49842381  &  H   \\
		-4.86810906   &    -1.87826311    &    3.41237907  &  H   \\
		-4.19804636   &     3.03099796    &   -0.34030993  &  O   \\
		-3.34494555   &     2.85880562    &    0.09007329  &  H   \\
		-4.67113657   &     2.18250698    &   -0.28466578  &  H   \\
		-10.07511676   &    -0.21674397    &    4.04062692  &  O   \\
		-9.10677318   &    -0.17829176    &    3.87801515  &  H   \\
		-10.15405972   &    -0.26984444    &    5.00795266  &  H   \\
		-6.77219247   &    -1.74224170    &   -0.14140135  &  O   \\
		-6.73602730   &    -2.32170710    &   -0.90820607  &  H   \\
		-7.73078010   &    -1.47180490    &   -0.05150937  &  H   \\
		-3.87753490   &     0.81526216    &    2.44905474  &  O   \\
		-4.45739095   &     0.67749418    &    1.66759341  &  H   \\
		-3.09372630   &     1.27775839    &    2.10598890  &  H   \\
		-7.45615599   &     0.02868439    &    3.47198015  &  O   \\
		-7.07673800   &    -0.79757065    &    3.10583397  &  H   \\
		-7.54180095   &     0.64319841    &    2.71353921  &  H   \\
		-5.18336270   &    -1.25944908    &    6.35638930  &  O   \\
		-4.72667465   &    -1.36172289    &    5.47709693  &  H   \\
		-4.57459651   &    -1.60441908    &    7.01556225  &  H   \\
		-2.03982546   &     2.64647231    &    1.46303288  &  O   \\
		-1.07994241   &     2.69286166    &    1.43477380  &  H   \\
		-2.33612221   &     3.37884511    &    2.05891950  &  H   \\
		-9.19386157   &     3.93355691    &    4.22790270  &  O   \\
		-9.46863763   &     4.81710656    &    4.49384642  &  H   \\
	\end{matrix}
\end{equation*}

\begin{equation*}
	\begin{matrix}
	      -8.36594712   &     4.06685706    &    3.71985687  &  H   \\
	-10.95714191   &     2.10529553    &    2.92113036  &  O   \\
	-10.70905294   &     1.28146654    &    3.38996164  &  H   \\
	-10.41154587   &     2.79065888    &    3.33901006  &  H   \\
	-6.38413371   &    -2.25080930    &    2.51681354  &  O   \\
	-6.99305244   &    -2.91824828    &    2.88539959  &  H   \\
	-6.49012062   &    -2.25408943    &    1.54791623  &  H   \\
	-9.25663082   &    -1.11723776    &    0.23166429  &  O   \\
	-9.63855094   &    -1.60553470    &    0.99487590  &  H   \\
	-9.61330730   &    -0.21788412    &    0.28870937  &  H   \\
	-8.69294618   &     2.27277857    &    6.54712575  &  O   \\
	-8.92767462   &     2.83039515    &    5.78873499  &  H   \\
	-7.80107544   &     1.94005147    &    6.32988751  &  H   \\
	-3.16693800   &     4.51898755    &    2.99119271  &  O   \\
	-3.79823816   &     4.93443541    &    2.38211883  &  H   \\
	-3.73125910   &     4.00666473    &    3.59533428  &  H   \\
	-4.98171833   &     2.68631866    &    4.04572542  &  O   \\
	-5.40364777   &     2.17576386    &    4.76828004  &  H   \\
	-4.62244433   &     1.99212778    &    3.44954234  &  H   \\
	-7.65002602   &     1.77855785    &    1.47357284  &  O   \\
	-7.45909348   &     2.64740760    &    1.88872014  &  H   \\
	-8.51473390   &     1.83650838    &    1.02495085  &  H   \\
	-6.91073647   &     3.97214175    &    2.77970279  &  O   \\
	-6.19020458   &     3.51145856    &    3.28100093  &  H   \\
	-6.43433146   &     4.52962255    &    2.13784168  &  H   \\	
	\end{matrix}
\end{equation*}

\newpage
\item \textbf{d} \\
\begin{equation*}
	\begin{matrix}
		    10.47809957    &   12.71846274    &  -11.97899901  &  O    \\
		10.22658342    &   11.89488486    &  -11.52970077  &  H    \\
		10.63625709    &   12.49532394    &  -12.91029371  &  H    \\
		14.09311777    &    9.12486699    &  -17.83882468  &  O    \\
		14.26174373    &    8.51196398    &  -18.56013848  &  H    \\
		13.10630803    &    9.20798153    &  -17.77040535  &  H    \\
		11.69453872    &    9.05046944    &  -12.03404201  &  O    \\
		12.36811207    &    9.70239422    &  -11.73499234  &  H    \\
		10.88336492    &    9.32340836    &  -11.57297533  &  H    \\
		14.62855857    &    9.10960050    &  -15.14736265  &  O    \\
		15.04774856    &    9.99768970    &  -15.03923400  &  H    \\
		14.56493375    &    8.97001955    &  -16.11036183  &  H    \\
		13.32650857    &   12.38957845    &   -9.32295825  &  O    \\
		12.90076051    &   13.20379791    &   -9.65514210  &  H    \\
		14.19725408    &   12.62292269    &   -8.95504684  &  H    \\
		14.43476965    &   11.93385919    &  -17.40490086  &  O    \\
		14.94282447    &   11.94997045    &  -16.57508141  &  H    \\
		14.49004863    &   11.01159577    &  -17.70012802  &  H    \\
		11.93221506    &    9.46054541    &  -14.67181434  &  O    \\
		12.89870640    &    9.36433235    &  -14.77818760  &  H    \\
		11.75816995    &    9.22359852    &  -13.73655079  &  H    \\
		13.68652459    &    8.31293402    &   -8.87335154  &  O    \\
		13.67033485    &    7.85216735    &   -9.75079694  &  H    \\
		13.67255193    &    7.60370134    &   -8.22239526  &  H    \\
		13.63683144    &   10.74820678    &  -11.40476492  &  O    \\
	\end{matrix}
\end{equation*}

\begin{equation*}
	\begin{matrix}
		      14.56221812    &   10.44172678    &  -11.49171232  &  H    \\
		13.60096143    &   11.34194525    &  -10.62443672  &  H    \\
		11.70118834    &   10.28785052    &   -8.59490226  &  O    \\
		12.32392522    &    9.54731504    &   -8.70030405  &  H    \\
		12.25965659    &   11.07677455    &   -8.71232606  &  H    \\
		16.61741640    &   13.54203067    &  -11.14983835  &  O    \\
		15.81349085    &   14.04490669    &  -11.40797636  &  H    \\
		16.43050485    &   13.18561540    &  -10.26683097  &  H    \\
		9.87985870    &   10.38924823    &  -10.48691704  &  O    \\
		9.00248313    &   10.24051141    &  -10.12259110  &  H    \\
		10.50869353    &   10.34382583    &   -9.71743547  &  H    \\
		13.62492826    &   12.52626433    &  -13.39178907  &  O    \\
		12.76367580    &   12.44578183    &  -13.85300840  &  H    \\
		13.57364288    &   11.88608854    &  -12.64794357  &  H    \\
		16.13883173    &    9.81173163    &  -11.73868728  &  O    \\
		16.73224890    &   10.43103936    &  -12.20556120  &  H    \\
		16.08795271    &    8.99566520    &  -12.26703525  &  H    \\
		15.52426847    &    7.56884827    &  -13.18783620  &  O    \\
		15.20780246    &    8.02398796    &  -14.00378175  &  H    \\
		16.10322154    &    6.86075037    &  -13.48563599  &  H    \\
		14.39098362    &   14.78594157    &  -12.05049633  &  O    \\
		14.08762084    &   14.06646642    &  -12.63558853  &  H    \\
		13.62496120    &   14.91209799    &  -11.47230151  &  H    \\
		13.47332430    &    7.11773242    &  -11.22943415  &  O    \\
		12.73455972    &    7.64766434    &  -11.58197315  &  H    \\
	\end{matrix}
\end{equation*}

\begin{equation*}
	\begin{matrix}
	     14.16968424    &    7.19162291    &  -11.90028752  &  H    \\
	17.49712477    &   11.85293197    &  -12.90949885  &  O    \\
	17.17398081    &   12.53221290    &  -12.24768878  &  H    \\
	18.44525267    &   11.99645319    &  -12.98317492  &  H    \\
	11.27167404    &   12.06655647    &  -14.59614680  &  O    \\
	11.30210752    &   12.37717355    &  -15.51650851  &  H    \\
	11.46875076    &   11.10520456    &  -14.66550762  &  H    \\
	15.94176198    &   12.44968401    &   -8.62262091  &  O    \\
	16.41565701    &   12.70294167    &   -7.82525818  &  H    \\
	15.99909733    &   11.45857710    &   -8.67591215  &  H    \\
	15.57146298    &   11.57615800    &  -14.88087092  &  O    \\
	16.37366583    &   11.72493833    &  -14.35151506  &  H    \\
	14.84687749    &   11.96299904    &  -14.32863771  &  H    \\
	11.80450214    &   12.38394407    &  -17.30332256  &  O    \\
	11.57520664    &   13.07195211    &  -17.93516066  &  H    \\
	12.79058342    &   12.31404328    &  -17.33189525  &  H    \\
	16.02843534    &    9.85364647    &   -8.98165862  &  O    \\
	16.20377252    &    9.75353651    &   -9.93403619  &  H    \\
	15.21200427    &    9.34110670    &   -8.84457852  &  H    \\
	11.97249104    &   14.37757249    &  -10.55777948  &  O    \\
	11.42184672    &   13.78744725    &  -11.13159879  &  H    \\
	11.34642591    &   14.97802443    &  -10.14219663  &  H    \\
	11.51355405    &    9.54533398    &  -17.41872688  &  O    \\
	11.42251862    &   10.50999168    &  -17.49156768  &  H    \\
	11.42433060    &    9.37419908    &  -16.46557003  &  H    \\	
	\end{matrix}
\end{equation*}

\newpage
\item \textbf{e} \\
\begin{equation*}
	\begin{matrix}
		-0.86725915    &    3.70975315    &    2.59411208  &  O  \\
		-0.40769972    &    2.93270089    &    2.96642655  &  H  \\
		-1.61634808    &    3.38449245    &    2.06079563  &  H  \\
		-2.80406051    &    5.79337884    &   -0.23995074  &  O  \\
		-1.92617019    &    5.80859896    &   -0.67984184  &  H  \\
		-2.74068262    &    6.47681739    &    0.45221079  &  H  \\
		4.26880190    &    6.67983351    &    0.63834423  &  O  \\
		5.00097865    &    7.27966323    &    0.46753456  &  H  \\
		3.85124988    &    6.51942855    &   -0.24285212  &  H  \\
		1.92201346    &    7.10472250    &    1.92433925  &  O  \\
		1.93151953    &    6.96883271    &    2.89113413  &  H  \\
		2.84415981    &    7.06371237    &    1.60796604  &  H  \\
		-0.38476086    &    5.60254337    &   -1.35791339  &  O  \\
		0.09173621    &    5.39813668    &   -0.51435600  &  H  \\
		-0.04099486    &    6.46239781    &   -1.66166820  &  H  \\
		2.62346955    &    3.17463001    &   -0.06723413  &  O  \\
		1.95861397    &    3.82852465    &    0.23798401  &  H  \\
		2.66995149    &    3.29121485    &   -1.02892376  &  H  \\
		0.78828302    &    4.91242216    &    0.86879847  &  O  \\
		0.20578115    &    4.46848620    &    1.52297577  &  H  \\
		1.20062830    &    5.67839359    &    1.32203236  &  H  \\
		-2.19956215    &    7.65150579    &    1.70672297  &  O  \\
		-1.80058942    &    7.13880137    &    2.46082298  &  H  \\
		-2.83780010    &    8.24390471    &    2.11673630  &  H  \\
		-0.94169003    &    1.77560140    &   -0.68117308  &  O  \\
	\end{matrix}
\end{equation*}

\begin{equation*}
	\begin{matrix}
	     -0.58130708    &    2.31872433    &   -1.40917991  &  H  \\
	-1.52366250    &    2.38580977    &   -0.18452714  &  H  \\
	-1.17515246    &    6.21541086    &    3.64938347  &  O  \\
	-1.14091107    &    5.28069774    &    3.37441308  &  H  \\
	-0.29109810    &    6.37434169    &    4.02004746  &  H  \\
	-2.73546883    &    3.29211573    &    0.72653963  &  O  \\
	-3.57112994    &    2.87099158    &    0.46142005  &  H  \\
	-2.79845616    &    4.22141569    &    0.41013648  &  H  \\
	0.16002687    &    8.71450774    &    0.58503968  &  O  \\
	0.80230731    &    8.22544192    &    1.13368873  &  H  \\
	-0.70641938    &    8.42600994    &    0.92002094  &  H  \\
	1.21719477    &    1.00040500    &    0.87932802  &  O  \\
	1.83785519    &    1.65933366    &    0.51437079  &  H  \\
	0.40948010    &    1.16691730    &    0.35696529  &  H  \\
	-0.18838950    &    3.40145867    &   -2.79664881  &  O  \\
	-0.99431735    &    3.30092220    &   -3.33463186  &  H  \\
	-0.30547420    &    4.25795899    &   -2.32498647  &  H  \\
	0.76270178    &    8.00662289    &   -1.84659875  &  O  \\
	0.50780643    &    8.36671577    &   -0.95161071  &  H  \\
	0.58583428    &    8.70406355    &   -2.48391302  &  H  \\
	-3.15221354    &    0.85428930    &   -2.13654606  &  O  \\
	-3.03645373    &    1.44228223    &   -2.89980232  &  H  \\
	-2.32538662    &    0.98256979    &   -1.63322665  &  H  \\
	2.49580315    &    3.96667598    &    3.92605449  &  O  \\
	3.16925107    &    4.00701627    &    3.21309511  &  H  \\	
	\end{matrix}
\end{equation*}

\begin{equation*}
	\begin{matrix}
		     1.96179260    &    3.17696430    &    3.74399693  &  H  \\
		-4.29546743    &    4.46569054    &   -2.20249181  &  O  \\
		-4.69744863    &    3.77405343    &   -1.65213944  &  H  \\
		-3.87935213    &    5.06928359    &   -1.56392304  &  H  \\
		-2.71904527    &    3.10777008    &   -3.86936421  &  O  \\
		-3.28834527    &    3.67601034    &   -3.29153023  &  H  \\
		-3.01401926    &    3.28133968    &   -4.76847355  &  H  \\
		0.82519551    &    1.76005575    &    3.33676137  &  O  \\
		0.78615069    &    1.00215427    &    3.92693861  &  H  \\
		0.99380525    &    1.39423363    &    2.42115044  &  H  \\
		2.45505046    &    3.80091129    &   -2.80686555  &  O  \\
		1.52397300    &    3.56251471    &   -3.00386658  &  H  \\
		2.98199227    &    3.43177351    &   -3.52177597  &  H  \\
		1.51561874    &    6.30403677    &    4.45701685  &  O  \\
		1.85826197    &    5.37703866    &    4.30463279  &  H  \\
		1.85913931    &    6.57108215    &    5.31444017  &  H  \\
		4.32773512    &    4.11098255    &    1.91270714  &  O  \\
		4.47824235    &    5.00599992    &    1.57371018  &  H  \\
		3.89114039    &    3.67311373    &    1.16091911  &  H  \\
		-4.87224006    &    2.17896092    &   -0.61515456  &  O  \\
		-5.64563646    &    1.66633852    &   -0.36345059  &  H  \\
		-4.31487890    &    1.58455404    &   -1.18118414  &  H  \\
		3.03740748    &    6.36084721    &   -1.69058605  &  O  \\
		2.80026455    &    5.52007563    &   -2.11157256  &  H  \\
		2.28056829    &    6.95489723    &   -1.82084674  &  H  \\
	\end{matrix}
\end{equation*}

\end{itemize}

%\cite{Ihrig/etal:2015,Ren/etal:2021}

\begin{table*} [ht]
	\caption{The MP2, RPA@PBE, (RPA+rSE)@PBE, PBE0, and PBE single point energies for H$_2$O molecule. The structures are optimized for different levels of theory and basis set.
  For RPA@PBE and (RPA+rSE)@PBE the frozen-core approximation is used for structure relaxation and single point energy calculations.}
	\label{tab:water_single_molecule_energy}
	\centering
	\scalebox{1.}{
	\begin{tabular}{lr} 
		\hline\hline 
		\multicolumn{1}{c}{\multirow{2}{*}{H$_{2}$O~~~~~~~~~~~~~~~~~~~~~~~~~~~~~}} & \multicolumn{1}{c}{\multirow{2}{*}{~~~~~~~~~~~~~~~~~~~~~~~~~~~~~~~~~~~~~~~~~~~~E (Hartree)}}  \\
		 \\ 
		\hline 
   \multicolumn{2}{c}{Own Geometries} \\ 
  MP2[aTZ]$^{a}$            &   $-$76.328992    \\
  PBE[aTZ]$^{b}$                  &   $-$76.380357    \\
  PBE[QZ]$^{b}$                   &   $-$76.383283    \\
  PBE[light]$^{b}$                &   $-$76.383342    \\
  PBE[tight]$^{b}$                &   $-$76.388204   \\
  PBE0[QZ]$^{b}$                  &   $-$76.383553     \\
  \multicolumn{2}{c}{Geometry optimization RPA@PBE[QZ]} \\
   RPA@PBE[QZ]$^{b}$       &   $-$76.497036 \\
   MP2[QZ]$^{b}$           &   $-$76.347596  \\
  (RPA+rSE)@PBE[QZ]$^{b}$   &   $-$76.505122        \\
   \multicolumn{2}{c}{Geometry optimization PBE[QZ]} \\
  RPA@PBE[QZ]$^{b}$       &   $-$76.496966 \\
  MP2[QZ]$^{b}$           &   $-$76.347420 \\
   \multicolumn{2}{c}{Geometry used for TTM2.1-F with ($R_{O-H}$) 0.9578 \AA, ($\theta$) 104.51$^{\circ}$}  \\
   RPA@PBE[QZ]$^{b}$    &   $-$76.496984     \\  
   MP2[QZ]$^{b}$        & $-$76.347643     \\
 %\multicolumn{2}{c}{Geomtry optimization B3LYP/6-311++G(2d,2p)} \\
 %RPA@PBE[QZVPP]$^{b}$ & $-$76.497653 \\
 %(RPA+rSE)@PBE[QZVPP]$^{b}$ & $-$76.506427 \\ 
 % RPA@PBE[5Z]$^{b}$            & $-$76.514399 \\
 % RPA@PBE[6Z]$^{b}$   & $-$76.521127  \\
 % (RPA+rSE)@PBE[5Z]$^{b}$ & $-$76.523278 \\
 % (RPA+rSE)@PBE[6Z]$^{b}$           & $-$76.530318 \\
 %  RPA@PBE[CBS(5Z,6Z)]$^{b}$           & $-$76.530439 \\
 %  (RPA+rSE)@PBE[CBS(5Z,6Z)]$^{b}$                & $-$76.540062 \\
		\hline\hline
  $^a$Ref~\citenum{Rakshit_2019} & $^{b}$This work (results obtained using FHI-aims)\\
	\end{tabular}}
 \end{table*}
 
%\begin{table*} [ht!]
%	\caption{  The RPA-based structure parameters are obtained from structure relaxation using cc-pVQZ basis set and frozen-core approximation is used.  Here the average values of these parameters are listed. }
%	\label{tab:average_struc_parameter}
%	\begin{tabular}{lcccccc} 
%		\hline\hline 
%		\multicolumn{1}{c}{\multirow{2}{*}{}} &
%		\multicolumn{1}{c}{\multirow{2}{*}{}} &
%		\multicolumn{1}{c}{\multirow{2}{*}{$ R_{O\text{\textemdash}O} \text{(\AA)}$}} &
%		\multicolumn{1}{c}{\multirow{2}{*}{$ R_{hb} \text{(\AA)}$}} & \multicolumn{1}{c}{\multirow{2}{*}{$ R_{O\text{\textemdash}H} \text{(\AA)}$}} & \multicolumn{1}{c}{\multirow{2}{*}{$ \phi \text{(deg)}$}} & \multicolumn{1}{c}{\multirow{2}{*}{$\theta \text{(deg)}$}}  
%		\\ 
%		& & & \\
%		\hline 	\\ [-0.5ex]
%		RPA@PBE & 25--a & 2.74 & 1.77 & 0.98 &  169.1   &   105.3   \\  
%		        & 25--b & 2.74 & 1.77 & 0.98 &  168.8   &   105.2   \\
%		    	& 25--c & 2.73 & 1.77 & 0.98 &  168.3   &   105.0   \\  
%	        	& 25--d & 2.73 & 1.77 & 0.98 &   168.1  &   105.2   \\ 
%	        	& 25--e & 2.73 & 1.77 & 0.98 &   168.3  &    105.3  \\
	
%		\hline\hline
%	\end{tabular}
% \end{table*} 

\bibliography{./CommonBib}	
	
\end{document}